\documentclass[manuscript,screen]{acmart}

\usepackage{graphicx}
\usepackage{multirow}
\usepackage{subfigure}
\usepackage{amssymb}
\usepackage{algorithm}
\usepackage{verbatim}
\usepackage{bm}     
\usepackage{enumitem}       
\usepackage{bbding}       

\usepackage{algpseudocode}

\AtBeginDocument{%
  \providecommand\BibTeX{{%
    \normalfont B\kern-0.5em{\scshape i\kern-0.25em b}\kern-0.8em\TeX}}}

\setcopyright{acmcopyright}
\copyrightyear{2018}
\acmYear{2018}
\acmDOI{XXXXXXX.XXXXXXX}

\acmConference[Conference acronym 'XX]{Make sure to enter the correct
  conference title from your rights confirmation emai}{June 03--05,
  2018}{Woodstock, NY}
\acmPrice{15.00}
\acmISBN{978-1-4503-XXXX-X/18/06}

\begin{document}

\title{Triple Sequence Learning for Cross-domain Recommendation}

\author{Haokai Ma$^{\dagger, \bullet}$}
\affiliation{\institution{School of Software, Shandong University}
\city{Jinan}
\country{China}}
\email{mahaokai@mail.sdu.edu.cn}

\author{Ruobing Xie$^{\dagger}$}
\affiliation{\institution{WeChat, Tencent}
\city{Beijing}
\country{China}}
\email{ruobingxie@tencent.com}

\author{Lei Meng$^*$}
\affiliation{\institution{School of Software, Shandong University; }
\institution{Shandong Research Institute of Industrial Technology}
\city{Jinan}
\country{China}}
\email{lmeng@sdu.edu.cn}

\author{Xin Chen}
\affiliation{\institution{WeChat, Tencent}
\city{Beijing}
\country{China}}
\email{andrewxchen@tencent.com}

\author{Xu Zhang}
\affiliation{\institution{WeChat, Tencent}
\city{Beijing}
\country{China}}
\email{xuonezhang@tencent.com}

\author{Leyu Lin}
\affiliation{\institution{WeChat, Tencent}
\city{Beijing}
\country{China}}
\email{goshawklin@tencent.com}

\author{Jie Zhou}
\affiliation{\institution{WeChat, Tencent}
\city{Beijing}
\country{China}}
\email{withtomzhou@tencent.com}

\thanks{$^{\dagger}$ indicates equal contributions.}
\thanks{$^{*}$ indicates corresponding author.}
\thanks{$^{\bullet}$ This work has been done when the author was at Tencent for internship.}

\renewcommand{\shortauthors}{Haokai Ma et al.}

\vspace{-0.1cm}
\begin{abstract}

Cross-domain recommendation (CDR) aims to leverage the correlation of users' behaviors in both the source and target domains to improve the user preference modeling in the target domain. Conventional CDR methods typically explore the dual-relations between the source and target domains' behaviors. However, this may ignore the informative mixed behaviors that naturally reflect the user's global preference. To address this issue, we present a novel framework, termed triple sequence learning for cross-domain recommendation (Tri-CDR), which jointly models the source, target, and mixed behavior sequences to highlight the global and target preference and precisely model the triple correlation in CDR. Specifically, Tri-CDR independently models the hidden representations for the triple behavior sequences and proposes a triple cross-domain attention (TCA) method to emphasize the informative knowledge related to both user's global and target-domain preference. To comprehensively explore the cross-domain correlations, we design a triple contrastive learning (TCL) strategy that simultaneously considers the coarse-grained similarities and fine-grained distinctions among the triple sequences, ensuring the alignment while preserving information diversity in multi-domain. We conduct extensive experiments and analyses on six cross-domain settings. The significant improvements of Tri-CDR with different sequential encoders verify its effectiveness and universality. The code will be released upon acceptance.

\end{abstract}

\vspace{-0.1cm}

\vspace{-0.1cm}

\received{30 May 2023}
\received[revised]{12 March 2009}
\received[accepted]{5 June 2009}

\maketitle

\begin{figure}[!htbp]
\centering
\includegraphics[width=0.90\columnwidth]{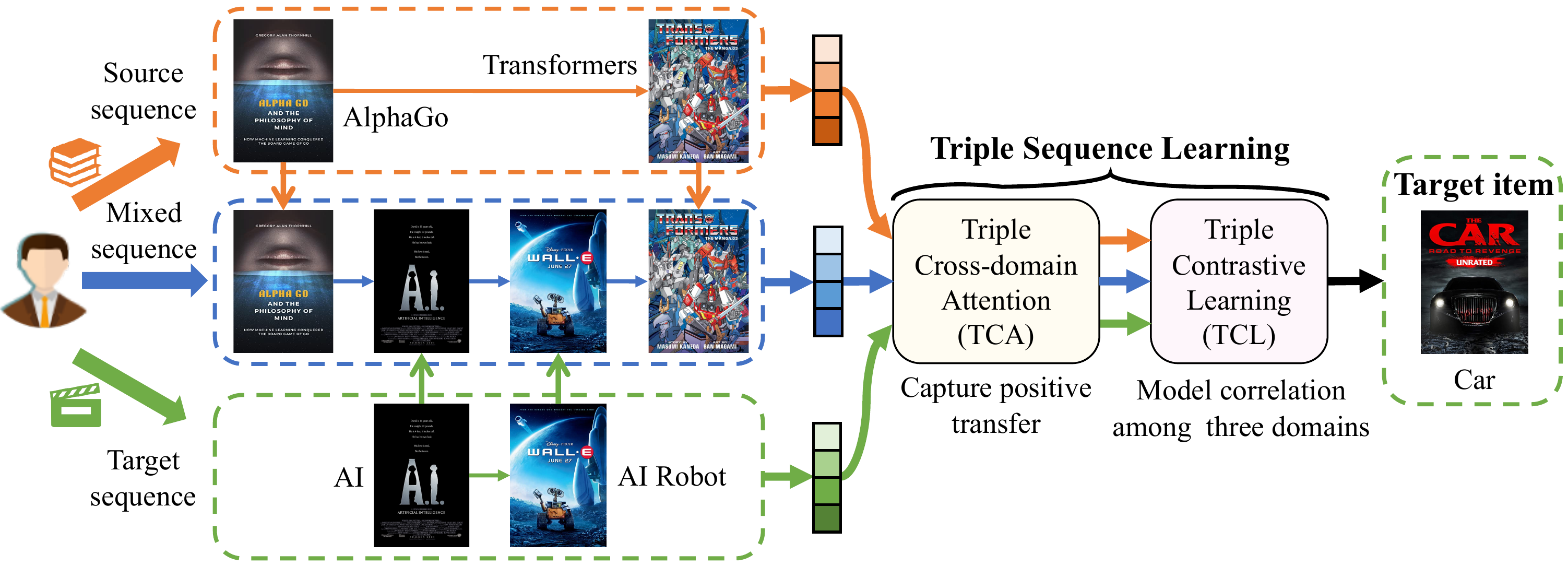}
\vspace{-0.2cm}
\caption{Triple sequence learning on the source, target, and mixed behavior sequences in CDR.}
\vspace{-0.4cm}
\label{fig:motivation_graph}
\end{figure}
\section{Introduction}
Personalized recommendation aims to capture user interests and provide appropriate items \cite{ADAPT, MEGCF}. Sequential recommendation (SR), which focuses on discovering user preferences from the essential information of users' historical behaviors, has attracted significant attention \cite{DIN,SASRec}. However, real-world SR models usually face the data sparsity problem, since users usually have few behaviors \cite{CL4SRec,MMCLR}. In practice, making full use of user behaviors in other domains under the user's approval is a straightforward and effective solution to the data sparsity issue in a single domain.

Cross-domain recommendation (CDR) concentrates on transferring useful information from the source domain to the target domain for performance gains in the target domain \cite{CoNET, DDTCDR, CDSRSurvey}.
Existing CDR methods mainly focus on modeling the relations between the source and target domains.
EMCDR \cite{EMCDR} and SSCDR \cite{SSCDR} attempt to learn a mapping function across the source/target domains via aligned objects. CoNet \cite{CoNET} and MiNet \cite{MiNet} adopt explicit cross-domain information paths or attention mechanisms for knowledge transfer. Some CDR methods further build the cross-domain connections via (multi-domain) global graphs \cite{CCDR, zhao2019cross} or feature correlations \cite{xie2020internal, DASL}. However, most existing CDR models simply focus on the dual relations between the source and target behavior sequences, ignoring the rich information of the natural mixed (i.e., source+target) behavior sequence.

We define the \textbf{mixed behavior sequence} in CDR as a complete user behavior sequence containing behaviors in both the source and target domains which are ordered chronologically. The left part of Figure \ref{fig:motivation_graph} shows an example of the source, target, and mixed behavior sequences in a cross-domain sequential recommendation (CDSR) scenario. The mixed sequence can reflect a user's complete behavioral pattern and preference evolution more comprehensively and thus helps to better extract users' global interests. For example, the sequential behaviors in domain \emph{book} and \emph{movie} are not consistent and reasonable. Only through the complete mixed sequence containing sequential behaviors of [\emph{book: AlphaGo}] $\rightarrow$ [\emph{movie: AI}] $\rightarrow$ [\emph{movie: AI Robot}] $\rightarrow$ [\emph{book: Transformers}] $\rightarrow$ [\emph{movie: Car}] can we fully understand the user's sequential action logic. We firmly believe that jointly modeling the mixed behavior sequence with the original two source/target sequences is beneficial to capture both inter-domain and intra-domain information in CDR.

In this work, we propose a new paradigm that jointly models source, target, and mixed behavior sequences in CDR. The challenges are three-fold:
(1) \textbf{\emph{How to extract more informative knowledge from source and mixed sequences?}}
User behaviors in other domains may be good supplements, while it is also common that users have different preferences in these domains. We should maximize the cross-domain information gain while alleviating the negative transfer from possible noises.
(2) \textbf{\emph{How to model the triple correlations among source, target, and mixed sequences?}}
The mixed sequence is built by source and target behaviors. Both the coarse-grained similarities and fine-grained distinctions among the three sequences should be carefully considered. The dual relation learning of conventional CDR models cannot be directly transferred to the triple learning task with the additional mixed sequence.
(3) \textbf{\emph{How to construct a universal CDR framework that could smoothly cooperate with different types of single-domain SR models?}}
Currently, lots of CDR models rely on complicated and customized networks for inter-domain interactions, which are hard to be directly adopted with other single-domain models. We aim to build a universal model-agnostic CDR framework that could be beneficial with frequently updated single-domain models.

To address these issues, we propose a novel \textbf{Triple sequence learning for cross-domain recommendation (Tri-CDR)}, serving as a model-agnostic framework to jointly model the source, target, and mixed sequences in CDSR. Specifically, we first build three sequence encoders for the source, mixed, and target domains, respectively, which model the intra-domain behavior interactions to get three hidden sequence representations.
Next, to alleviate the irrelevant negative transfer, we design a \textbf{triple cross-domain attention (TCA)} method on three sequence representations to capture the informative knowledge related to users' target-domain preferences and global interests. These attention-enhanced sequence representations are then combined and fed into a Multi-Layer Perceptron (MLP) to get the final user representation.
We further propose a \textbf{triple contrastive learning (TCL)} strategy to comprehensively model the correlations among three sequences. TCL adopts three CL losses to capture the coarse-grained similarities between any two sequence representations of the same user compared to other users'. More importantly, it further employs a margin-based triple loss among three sequence representations to model their fine-grained distinctions, keeping the information diversity in three domains.
The advantages of Tri-CDR are: (1) the TCA enables an informative knowledge transfer related to users' target-domain preferences and global interests. (2) The TCL helps to better capture the correlations among three domains in representation learning. (3) Tri-CDR is effective, universal, and easy-to-deploy, which could be conveniently applied with different sequence encoders and additional objectives.

In experiments, we have conducted an extensive evaluation on six cross-domain settings with various base sequence encoders. As a result, Tri-CDR achieves significant improvements on all settings. We also conduct various ablation studies, universality analyses, parameter analyses, and visualization to verify the effectiveness of the proposed TCA and TCL. The contributions are summarized as follows:

\begin{itemize}[leftmargin=*, topsep=0.2pt,parsep=0pt]
  \item We have verified the significance of the triple sequence modeling for comprehensive user interest understanding. To the best of our knowledge, we are the first to present the triple sequence learning among source, target, and mixed behavior sequences in CDR.
  \item We propose a triple cross-domain attention (TCA) method to enable more positive transfer of knowledge from the source domain to the target one, which considers both the user's target-domain preferences and global interest from the source and mixed behavior sequences.
  \item We creatively design the triple contrastive learning (TCL) strategy, which not only models the coarse-grained similarities among multi-domain sequence representations of the same user but also detects the fine-grained distinctions via a margin-based triple loss.
  \item We conduct an extensive evaluation to verify the effectiveness of our Tri-CDR on multiple datasets with different base models. The proposed model is effective, universal, and easy-to-deploy.
\end{itemize}

\section{Related Work}
\textbf{Cross-domain Recommendation.}
Cross-domain recommendation (CDR) is a representative method to alleviate the data sparsity problem in a single domain with auxiliary information from other domains \cite{EMCDR, DDTCDR}.
The basic assumption is that users' behaviors in different domains reflect the user's personal preferences to a certain extent.
Classical CDR methods aim to model cross-domain knowledge transfer through directly mapping \cite{EMCDR, SSCDR}, multi-domain interaction modeling \cite{MiNet,HCDIR}, meta-learning \cite{TMCDR, PTUPCDR}, and transformation matrix \cite{CoNET, DDTCDR}. Recently, some CDR methods also leverage alignment constraint \cite{LACDR, DisAlign} and adversarial learning \cite{RecGURU} for cross-domain knowledge representation and fusion. CDSR concentrates more on users' multi-domain sequential behavior modeling in CDR \cite{BiTGCF,piNet,DDGHM,DASL}.
BiTGCF \cite{BiTGCF} designs a bi-directional transfer learning method to transfer users' single-domain preferences across two domains.
Lots of CDSR methods focus on the shared-account CDSR scenario \cite{piNet,CDHRM}. $\pi$-Net \cite{piNet} and PSJNet \cite{PSJNet} utilize a shared account filter unit and a cross-domain transfer unit to share information across both domains synchronously. MIFN \cite{MIFN} further enriches the sequence representation with knowledge graphs based on $\pi$-Net. DDGHM \cite{DDGHM} builds a global dynamic graph to model source-target interactions directly, and jointly predicts via local and global information.
CDHRM \cite{CDHRM} jointly captures users' inter-session and intra-session behavioral dynamics from different domains. DAGCN \cite{DAGCN} designs a domain-aware graph convolution network to learn user-specific node representations on the global static graph. 
DASL \cite{DASL} proposes the dual embedding and dual attention strategies to model the correlations between source and target domains' sequences.
DR-MTCDR \cite{DR-MTCDR} designs a unified disentanglement module to capture the domain-shared and domain-specific information, with the aim to transfer the trustworthy information across domains.
COAST \cite{COAST} introduces a unified cross-domain heterogeneous graph to capture user-item similarity and achieve user interest alignment by exploring the user-user and user-item interest invariance.
However, existing CDSR methods merely focus on dual relations of source$\rightarrow$target or global$\rightarrow$local, ignoring directly modeling the natural mixed behavior sequence with the correlations among source, target, and mixed domains. Our Tri-CDR is different from these works: (a) we directly model three mixed, source, and target behavior sequences and use them for recommendations simultaneously. (b) We emphasize the ternary relationship among three sequences for positive knowledge transfer. (c) Tri-CDR is model-agnostic and easy-to-deploy that can be applied to different base sequence encoders and even intra-domain CL tasks.
By doing this, our work can not only learn both complete and independent preferences from mixed, source, and target behavior sequences but also explain users' sequential behavior comprehensively through their interactions.

\noindent
\textbf{Sequential Recommendation.}
Sequential Recommendation (SR) attempts to capture the user's time-aware preferences by modeling the sequential dependencies of the user's historical behavior to recommend the next item that the user may be interested in. Early works reason users' short-term preference through the Markov Chains (MCs) \cite{MC1, MC2}. In recent years, researchers have leveraged the Convolution Neural Network (CNN) \cite{Caser, RCNN}, Recurrent Neural Network (RNN) \cite{GRU4Rec, NARM} and Transformer \cite{SASRec, BERT4Rec} to capture users' preference patterns from users' historical behaviors. Among them, GRU4Rec \cite{GRU4Rec} leverages the Gate Recurrent Unit (GRU) as the sequential encoder to learn users' long-term dependencies. SASRec \cite{SASRec} introduces Transformer for behavior interaction modeling and is widely used in practice. S$^3$Rec \cite{zhou2020s3} leverage the mutual information maximization principle to employ four self-supervised objectives among item, attribute, sub-sequence, and sequence. DUVRec \cite{DUVRec} encodes the sequential information with dual-view user representation (item-view and factor-view) to achieve enhanced SR performance. CL4SRec \cite{CL4SRec} is one of the state-of-the-art SR models that further enhance the sequential modeling with various intra-domain CL tasks. In this work, we have successfully adopted Tri-CDR with different sequence encoders, including GRU4Rec, SASRec, and CL4SRec.

\noindent
\textbf{Contrastive Learning in Recommendation.}
As a common self-supervised learning (SSL) method, contrastive learning (CL) has been widely used in fields of Computer Vision (CV) \cite{Moco, SimCLR}, Natural Language Processing (NLP) \cite{DeCLUTR, CLEAR} and Recommendation System (RS) \cite{CLCRec, CL4SRec, CCDR}. In recommendation, CL is widely applied to session-based recommendation \cite{C2CRS}, multi-behavior recommendation \cite{MMCLR}, sequential recommendation \cite{DUORec,zhou2020s3, CL4SRec}, and cross-domain recommendation \cite{CCDR, C2CDR}. C$^2$-CRS \cite{C2CRS} proposes a coarse-to-fine contrastive learning method to model user preference with multi-level semantic fusion. MMCLR \cite{MMCLR} designs three CL tasks to learn the correlations among different behavior types and modeling views. DUORec \cite{DUORec} proposes a contrastive regularization with the model-level augmentation to reshape and improve the embedding distribution and sequence representations. CL4SRec \cite{CL4SRec} proposes three sequence-based augmentations to build positive pairs in SSL. CCDR \cite{CCDR} designs an intra-domain CL task and three inter-domain CL tasks for cross-domain knowledge transfer in graph-based matching. C$^{2}$DSR \cite{C2CDR} conducts a cross-domain infomax objective to enhance the correlation between global and local representations with domain-specific global augmentations. Some recent works \cite{TIL, CauseRec, TLAutopool, MMCLR} have conducted certain triplet losses for more precise representation learning in recommendation. However, existing CL-based CDR models simply maximize the mutual information of representations in different domains, ignoring their conflicts that may lead to negative transfer and model collapse.
To the best of our knowledge, we are the first to jointly model coarse-grained similarity and fine-grained distinction via CL in CDR.

\section{Method}
\subsection{Problem Formulation}

We first define the \emph{source behavior sequence} $S^{S}=\{\bm{d}^S_1, \bm{d}^S_2, \cdots, \bm{d}^S_p\}$ in the source domain $S$ and the \emph{target behavior sequence} $S^{T}=\{\bm{d}^T_1, \bm{d}^T_2, \cdots, \bm{d}^T_q\}$ in the target domain $T$ for each user, where $p$, $q$ are the source/target historical behavior lengths, and $\bm{d}^S_i$ and $\bm{d}^T_j$ are behavior embeddings. In Tri-CDR, we propose a third \emph{mixed behavior sequence} $S^{M}=\{\bm{d}^M_1, \bm{d}^M_2, \cdots, \bm{d}^M_{p+q}\}$ as a supplement to source/target sequences, which is the complete user behavior sequence containing both source and target behaviors in chronological order. Given three behavior sequences $S^{S}$, $S^{T}$ and $S^{M}$, Tri-CDR tries to recommend the target item $d^T_{q+1}$ that will be interacted by this user in the target domain.

\begin{figure*}[!t]
\centering
\includegraphics[width=0.98\textwidth]{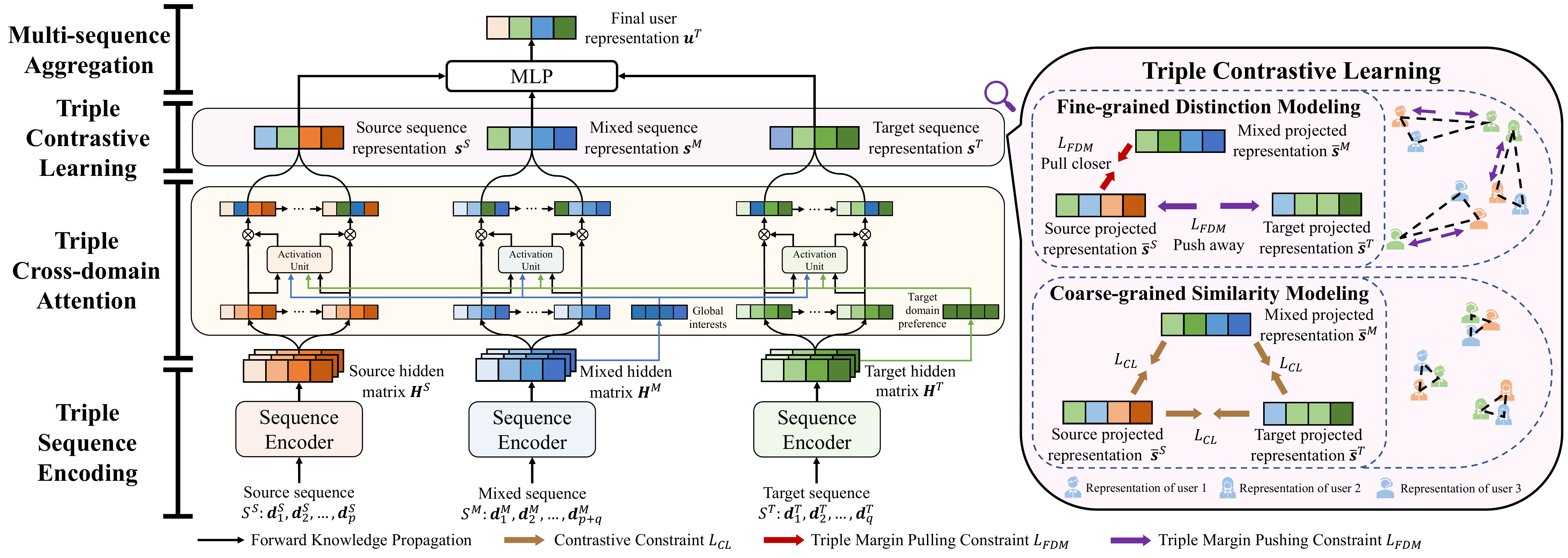}
\vspace{-0.2cm}
\caption{The overall structure of Tri-CDR. TCA highlights informative knowledge related to users' target-domain preferences and global interests, while TCL captures the triple correlations among three domains for better cross-domain knowledge transfer.}
\vspace{-0.4cm}
\label{fig:overall_structure}
\end{figure*}


\subsection{Overall Framework}
In this section, we describe the proposed model-agnostic Triple sequence learning for cross-domain recommendation (Tri-CDR) framework, which jointly models source, target, and mixed behavior sequences to improve CDR. Specifically, we first model the source, target, and mixed behavior sequences through three base sequence encoders separately to generate their corresponding hidden sequence representations in three domains. Then we propose a triple cross-domain attention (TCA) method to highlight informative knowledge related to the user's target-domain preferences and global interests in building three domains' sequence representations, mitigating negative knowledge transfer. To better model the correlations among multi-domain sequence representations, we design a novel triple contrastive learning (TCL) strategy with two contrastive constraints: (a) Coarse-grained similarity modeling, which enables source/target/mixed sequence representations from the same user to be more similar than other users'. (b) Fine-grained distinction modeling, which recognizes users' diversified preferences in different domains to keep the information gains brought by the source and mixed sequences. TCL helps to learn more informative and accurate multi-domain representations to capture user preferences comprehensively.

The overall structure of Tri-CDR is illustrated in Fig. \ref{fig:overall_structure}. In the following subsections, we first present the base sequence encoder of the proposed Tri-CDR which is implemented with the self-attention module. Subsequently, we introduce TCA, which emphasizes the user's target-domain preferences and global interest in triple domains. And then, we describe TCL to precisely model the user's coarse-grained similarities and fine-grained distinctions among triple sequences. Finally, we present the discussions on the proposed Tri-CDR.

\subsection{Base Sequence Encoder}

Inspired by the success of the self-attention mechanism in sequential recommendation, we apply SASRec \cite{SASRec} as our sequence encoder for all domains. Without losing generality, for the target domain sequence $S^T$, we build the input matrix $\bm{D}^T \in \mathbb{R}^{q \times d}$, where each behavior embedding $\bm{d}^T_i$ consists of a learnable item ID embedding and a position embedding, and $d$ is the embedding size. Then the sequence encoder transposes $\bm{D}^T$ into three matrices by linear projections, and feeds them into the attention method as query, key, and value, which can be defined as:
\begin{equation}
    \bm{\hat{H}}^T = \operatorname{Attention}(\bm{Q}, \bm{K}, \bm{V})=\operatorname{Softmax}\left(\frac{\bm{Q K}^T}{\sqrt{d}}\right) \bm{V}.
\end{equation}
where $ \bm{Q}= \bm{D}^T \bm{W}^Q, \bm{K}= \bm{D}^T \bm{W}^K, \bm{V}= \bm{D}^T \bm{W}^V$, and $\bm{W}^Q, \bm{W}^K, \bm{W}^V$ denote the linear projections respectively. Meanwhile, we conduct a point-wise feed-forward network to get the hidden behavior matrix $\bm{H}^T$ of the target domain, which is measured as follows:
\begin{equation}
    \bm{H}^T=\operatorname{ReLU}\left(\bm{\hat{H}}^T \bm{w}_1+\bm{b}_1\right) \bm{w}_2+\bm{b}_2, \quad \bm{H}^T \in \mathbb{R}^{q \times d}.
\end{equation}
where $\bm{w}_1$, $\bm{w}_2$, $\bm{b}_1$, $\bm{b}_2$ denote the weight matrices and bias vectors respectively. The hidden matrices of source/mixed sequences $\bm{H}^S$, $\bm{H}^M$ are similarly constructed with the intra-domain behavior interactions. Note that the same items in different domains are allocated with different behavior embeddings for better representation capacity, avoiding too homogeneous representations across different domains. It is also worth noting that we can conveniently adopt other sequential models as our sequence encoder or even with other intra-domain CL tasks in Tri-CDR (we have tested GRU4Rec \cite{GRU4Rec} and CL4SRec \cite{CL4SRec}, see the universality analysis in Sec. \ref{sec.universality}).

\subsection{Triple Cross-domain Attention}
\label{par:TCA}

We assume that different historical behaviors in three domains should have different importance for the target-domain prediction. Precisely, We expect that the information emphasized in three sequences should be (a) relevant to the user's target-domain preferences, so as to fit the target-domain prediction task, and (b) relevant to the user's global interests, so as to understand the user's comprehensive preferences and bring in more information gain. Hence, we propose a \emph{Triple Cross-domain Attention (TCA)} on three hidden behavior matrices to enable more positive transfer.

TCA functions when we aggregate the hidden behavior embeddings in $\bm{H}^S$, $\bm{H}^T$, $\bm{H}^M$ to get three sequence representations $\bm{s}^S$, $\bm{s}^T$, $\bm{s}^M$ of the source, target, and mixed domains.
Specifically, for the source domain, given $\bm{H}^S=\{\bm{h}_1^S, \cdots, \bm{h}_p^S\}$ and the target item $d^T_{q+1}$, TCA calculates the attention weight $\alpha_i^S$ of the $i$-th behavior's hidden embedding $\bm{h}_i^S$ in the source sequence as follows:
\begin{equation}
    \alpha_i^S=\operatorname{f}(\bm{h}_q^T,\bm{h}_{p+q}^M,\bm{h}_i^S) = \operatorname{MLP}^a\!\left(\bm{h}_q^T \| \bm{h}_i^S \| \bm{h}_i^S\!-\!\bm{h}_q^T \| \bm{h}_i^S\!\odot\!\bm{h}_q^T \| \bm{h}_{p+q}^M \right).
\end{equation}
Here, $\odot$ performs element-wise vector product, $\|$ denotes the concatenation operation, and $\operatorname{MLP}^a(\cdot)$ is a two-layer fully-connected network where each layer is followed by a PReLU activation function. $\bm{h}_q^T$ and $\bm{h}_{p+q}^M$ denote the last hidden behavior embeddings in target and mixed domains. 
This attention setting is inspired by \cite{DIN}, while we adopt $\bm{h}_q^T$ to indicate the user's current target-domain preference instead of using different candidate target item's embedding $\bm{d}^T_{q+1}$, since it is much more efficient in online serving. Moreover, we also highlight $\bm{h}_{p+q}^M$ in this attention, since we assume the hidden embedding of the last mixed behavior implies the user's global interests, which are good supplements to the target-domain preference.
Based on this, we aggregate different behavior hidden embeddings $\bm{h}_i^S$ to get the source sequence representation $\bm{s}^S$ as:
\begin{equation}
    \bm{s}^S = \sum_{i=1}^p \operatorname{Softmax} \left( \alpha_i^S \right) \bm{h}_i^S.
\end{equation}
Similarly, we also have $\bm{s}^M\!=\!\sum_{i=1}^{p+q}\!\operatorname{Softmax}\!\left( \operatorname{f}(\bm{h}_q^T,\bm{h}_{p+q}^M,\bm{h}_i^M) \right) \bm{h}_i^M$ and $\bm{s}^T\!=\!\sum_{i=1}^{q}\!\operatorname{Softmax}\!\left( \operatorname{f}(\bm{h}_q^T,\bm{h}_{p+q}^M,\bm{h}_i^T) \right) \bm{h}_i^T$ for the mixed and target sequence representations. With TCA, we can not only directly focus on the information related to the user's target-domain preferences, but also keep aware of the user's global interests for a more comprehensive positive transfer from the three domains.

\subsection{Triple Contrastive Learning}
\label{par:TCL}

Compared to classical CDR models that learn dual relations, Tri-CDR faces a more challenging task to comprehensively understand triple correlations among source, target, and mixed sequences. In this work, we propose a novel \emph{Triple Contrastive Learning (TCL)} to smartly model the correlations among three sequence representations. Precisely, we design two CL tasks in TCL, including the coarse-grained similarity modeling and the fine-grained distinction modeling. The former CL task is adopted to capture the coarse-grained similarities between any two sequence representations of the same user compared to others. In contrast, the latter CL task is conducted to model the fine-grained distinctions among a user's multi-domain sequence representations, keeping the diversity across different domains to enhance information gains.

\subsubsection{Coarse-grained Similarity Modeling (CSM)}

It is natural that a user's behavior sequences in different domains should share common general preferences. Hence, we design a \textbf{coarse-grained similarity modeling (CSM)} to model the coarse-grained similarities among three domains' sequence representations of the same user. Specifically, we project the sequence representations $\bm{s}^S$, $\bm{s}^T$, $\bm{s}^M$ into their spaces via domain-specific projectors $\operatorname{P}^S(\cdot)$, $\operatorname{P}^T(\cdot)$, $\operatorname{P}^M(\cdot)$ (we build these projectors via one-layer MLPs). After obtaining the projected sequence representations $\bm{\bar{s}}^S=\operatorname{P}^S(\bm{s}^S)$, $\bm{\bar{s}}^T=\operatorname{P}^T(\bm{s}^T)$, $\bm{\bar{s}}^M=\operatorname{P}^M(\bm{s}^M)$ of three domains, we calculate the contrastive loss $\mathcal{L}_{CL}$ with any two of them as positive instances in CL. Formally, for $\bm{\bar{s}}^S$ and $\bm{\bar{s}}^T$, we follow the classical InfoNCE \cite{SimCLR} as:
\begin{equation}
    \label{infonce}
    \mathcal{L}_{CL}\left(\bm{\bar{s}}^S,\bm{\bar{s}}^T\right)=-\sum_{i \in B} \log \frac{\operatorname{exp}(\operatorname{sim}\left(\bm{\bar{s}}_i^S, \bm{\bar{s}}_i^T\right) / \tau)}{\sum_{j \in B \setminus i} \operatorname{exp} (\operatorname{sim} \left(\bm{\bar{s}}_i^S, \bm{\bar{s}}_j^T\right) / \tau)}.
\end{equation}
$B$ denotes the sampled batch, $\bm{\bar{s}}_j^T$ denotes the randomly selected negative sample for $i$ in $B$, $\tau$ denotes the temperature coefficient, and $\operatorname{sim}\left(\bm{\bar{s}}_i^S, \bm{\bar{s}}_i^T\right) \!=\!\frac{{(\bm{\bar{s}}_i^S)}^{T} (\bm{\bar{s}}_i^T)}{\|\bm{\bar{s}}_i^S\| \cdot\left\|\bm{\bar{s}}_i^T\right\|}$ denotes the cosine similarity. Finally, the CSM loss $\mathcal{L}_{CSM}$ is formulated with three positive pairs as:
\begin{equation}
    \label{eq.l_csm}
    \mathcal{L}_{CSM}\!=\!\sum_{u}(\lambda_{1} \mathcal{L}_{CL}\!\left(\bm{\bar{s}}^M,\bm{\bar{s}}^S\right)+\lambda_{2} \mathcal{L}_{CL}\!\left(\bm{\bar{s}}^M,\bm{\bar{s}}^T\right)+\lambda_{3} \mathcal{L}_{CL}\!\left(\bm{\bar{s}}^S,\bm{\bar{s}}^T\right)), \\
\end{equation}
where $\lambda_1$, $\lambda_2$, and $\lambda_3$ denote the loss weights respectively.

\subsubsection{Fine-grained Distinction Modeling (FDM)}

CSM assumes that a user's domain-specific preferences should be more similar, and monotonously pulls the multi-domain representations of the same user closer. It functions well in general, while there does exist fine-grained distinctions across the user's preferences in triple domains. Over-optimizing $\mathcal{L}_{CSM}$ may inevitably cause the model to collapse to the proximal point, where the source and mixed sequences cannot provide additional information gain for their too-similar target sequence, putting the cart before the horse.

To address this issue, we look back to the composition of the proposed mixed sequence, whose subsequences contain both source and target sequences. Hence, it is intuitive that the source-mixed and target-mixed distances should be smaller than the source-target distance. Under this intuition, we propose a new CL task named \textbf{fine-grained distinction modeling (FDM)} based on a margin-based triplet loss. We intuitively assume the distance between $\bm{\bar{s}}^S$ and $\bm{\bar{s}}^T$ should be larger than the distance between $\bm{\bar{s}}^S$ and $\bm{\bar{s}}^M$ by at least the margin $\gamma$. Formally, we define the FDM loss $\mathcal{L}_{FDM}$ as:
\begin{equation}
    \mathcal{L}_{FDM}=\sum_{u} \max \left\{d\left(\bm{\bar{s}}^S,\bm{\bar{s}}^M\right)-d\left(\bm{\bar{s}}^S, \bm{\bar{s}}^T\right)+\gamma, 0\right\},
\label{eq.L_FD}
\end{equation}
where $d\!\left(\!\bm{\bar{s}}^S\!,\!\bm{\bar{s}}^T\!\right)$ measures the similarity between $\bm{\bar{s}}^S$ and $\bm{\bar{s}}^T$, which is calculated as the $L_2$ distance. $\gamma$ denotes the margin to be controlled.
Through jointly optimizing $\mathcal{L}_{CSM}$ and $\mathbf{L}_{FDM}$, Tri-CDR could intelligently find a good balance in ensuring the alignment among a user's multi-domain preferences while keeping the fine-grained distinctions to bring in more information gain from other domains.

\subsection{Optimization Objectives}

After TCA and TCL, we concatenate the triple sequence representations $\bm{s}^M$, $\bm{s}^S$, $\bm{s}^T$ and feed them into the following multi-sequence aggregation layer to generate the user final representation $\bm{u}^T$, which is formulated as follows:
\begin{equation}
    \bm{u}^T=\operatorname{MLP}^f(\bm{s}^M \ \| \ \bm{s}^S \ \| \ \bm{s}^T).
\end{equation}
$\operatorname{MLP}^f(\cdot)$ denotes a two-layer fully-connected network with the $\operatorname{LeakyReLU}$ activation. Finally, we calculate the predicted probability $\hat{y}^T=(\bm{u}^T)^\top \bm{d}_{q+1}^T$ of the user $u$ on the target item $d_{q+1}^T$ with the final user representation $\bm{u}^T$ and the item embedding $\bm{d}_{q+1}^T$. We formulate the binary cross-entropy loss $\mathcal{L}_{CTR}$ as follows:
\begin{equation}
    \mathcal{L}_{CTR}\!=\!-\!\sum_{(u,d) \in R^T} \left[y_{u,d}^T \log \hat{y}_{u, d}^T\!+\!\left(1\!-\!y_{u,d}^T\right) \log \left(1\!-\!\hat{y}_{u,d}^T\right)\right]  
    \label{eq.L_CTR}
\end{equation}
where $R^T$ is the target-domain training set, $y_{u,d}^T\!=\!1$ and $y_{u,d}^T\!=\!0$ denote the positive and negative samples respectively, and $\hat{y}_{u,d}^T$ denotes the predicted probability of $(u,d)$. To optimize across triple sequences in conjunction with CL tasks, the overall objective function $L$ is a linear combination of $\mathcal{L}_{CTR}$, $\mathcal{L}_{CSM}$ and $\mathcal{L}_{FDM}$ as:
\begin{equation}
    \mathcal{L}=\mathcal{L}_{CTR}+\lambda_{CSM}\mathcal{L}_{CSM}+\lambda_{FDM}\mathcal{L}_{FDM}
\label{eq.L_all}
\end{equation}
where $\lambda_{CSM}$ and $\lambda_{FDM}$ denote the loss weights of $\mathcal{L}_{CSM}$ and $\mathcal{L}_{FDM}$. 

\subsection{Training Strategy of CDSR}
\label{app.implementation}

Training Tri-CDR from scratch sometimes may incur the difficulty in model convergence in the early training stage, leading to unstable and unsatisfactory performance. This is mainly caused by the feature space conflicts between the source and target domains in CDR. We also observe the same phenomenon in other CDSR models. 
To address this issue, we conduct a two-step training strategy (i.e., single-domain pre-training + cross-domain fine-tuning) to achieve more stable and satisfactory performances. Specifically, we first pre-train two source/target SASRec models with their corresponding single-domain losses. Next, the pre-trained SASRec parameters are used as the initialization of Tri-CDR (and other CDR baselines). All parameters are then tuned via $L$ in Eq. (\ref{eq.L_all}). Through this, CDR models could be trained more effectively and stably. 

\vspace{-0.2cm}
\subsection{In-depth Model Discussions}
In this section, we undertake a comparison between the proposed Tri-CDR and the existing CDSR methods specifically tailored for the domain utilization, CL implementation, sequence modeling universality and complexity analyses, intending to analyze its novelty and effectiveness. 

\begin{table}[t]
\caption{Comparison among representative cross-domain sequential recommendation algorithms.}
\vspace{-0.3cm}
\label{tab:method}
\small
\begin{tabular}{|c|ccc|c|c|}
\hline
\multirow{2}{*}{\textbf{Algorithms}} &
  \multicolumn{3}{c|}{Domain utilization} &
  \multirow{2}{*}{\begin{tabular}[c]{@{}c@{}}Contrastive Learning \\ Implementation\end{tabular}} &
  \multirow{2}{*}{\begin{tabular}[c]{@{}c@{}}Sequence modeling \\ universality\end{tabular}} \\ \cline{2-4}
       & \multicolumn{1}{c|}{Mixed Domain} & \multicolumn{1}{c|}{Source Domain} & Target Domain &          &          \\ \hline
$\pi$-Net \cite{piNet}  & \multicolumn{1}{c|}{$\times$}     & \multicolumn{1}{c|}{$\checkmark$}  & $\checkmark$  & $\times$ & $\times$ \\ \hline
DTCDR \cite{DTCDR}  & \multicolumn{1}{c|}{$\times$}     & \multicolumn{1}{c|}{$\checkmark$}  & $\checkmark$  & $\times$ & $\times$ \\ \hline
DDTCDR \cite{DDTCDR} & \multicolumn{1}{c|}{$\times$}     & \multicolumn{1}{c|}{$\checkmark$}  & $\checkmark$  & $\times$ & $\times$ \\ \hline
DASL \cite{DASL}   & \multicolumn{1}{c|}{$\times$}     & \multicolumn{1}{c|}{$\checkmark$}  & $\checkmark$  & $\times$ & $\times$ \\ \hline
DDGHM \cite{DDGHM}  & \multicolumn{1}{c|}{$\checkmark$} & \multicolumn{1}{c|}{$\times$}      & $\checkmark$  & intra-CL & $\times$ \\ \hline
C$^2$DSR \cite{C2CDR}  & \multicolumn{1}{c|}{$\checkmark$} & \multicolumn{1}{c|}{$\times$}      & $\checkmark$  & inter-CL & $\times$ \\ \hline
Tri-CDR (ours) &
  \multicolumn{1}{c|}{$\checkmark$} &
  \multicolumn{1}{c|}{$\checkmark$} &
  $\checkmark$ &
  Both intra- and inter- CL &
  $\checkmark$ \\ \hline
\end{tabular}
\vspace{-0.4cm}
\end{table}

\subsubsection{\textbf{Comparison with Existing CDR Methods in Domain Utilization.}} 
CDSR aims to predict the next item that the user will be consumed in the target domain by leveraging the historical behavior sequence in the source domain. Therefore, the pioneer CDSR methods primarily focus on exploring how to achieve meaningful information transfer with the leverage of user's behavior in the source domain \cite{DTCDR,DDTCDR,DASL}. 
With the profound advancement of CDSR, some studies attempt to incorporate the mixed behavior sequence containing both source and target behaviors in chronological order to model the user's global interests. However, these studies are specifically designed for Cross-domain Session-based Recommendation \cite{CDHRM} and Cross-domain Share-account recommendation \cite{piNet}, and may not be directly applicable to CDSR scenarios. 
Recently, DDGHM \cite{DDGHM} and C$^2$DSR \cite{C2CDR} introduce the mixed behavior sequence into CDSR through the utilization of the global graph, yielding promising performance. Nevertheless, the aforementioned approaches typically model users' global and local interests separately with the mixed and target behavior sequences to capture their dynamic correlations. Despite the commendable performance of these methods, it overlooks the modeling of the user's preference in the source domain, which is the coherent sequence with its internal logic and equally critical for CDSR. Consequently, it may result in sub-optimal performance.

In contrast, the proposed Tri-CDR jointly models the user's dynamic preference from the source, mixed and target domain, which facilitates the flexible and convenient integration of all available information into CDSR scenarios. Meanwhile, Tri-CDR proposes the Triple Cross-domain Attention mechanisms and Triple Contrastive Learning strategies for negative filtering, enabling fine-grained inter-domain relationship modeling and precise cross-domain positive knowledge transfer. The source, mixed and target domains are beneficial in CDR, while it is non-trivial to make full use of them.
More detailed experimental results and analyses of the domain utilization are described in Sec.\ref{sec.domain_analysis}. 

\subsubsection{\textbf{Comparison with Existing CDR Methods in CL Implementation.}}
As a self-supervised learning strategy, CL has been widely applied to collaborative filtering \cite{NCL}, multi-media recommendation \cite{CLCRec}, and sequential recommendation \cite{CL4SRec, DUORec}. Recently, some studies \cite{DDGHM, C2CDR, CCDR} have also introduced CL into CDR. DDGHM \cite{DDGHM} designs an intra-domain contrastive metric with the random item augmentation operator to enhance representation learning and alleviate data sparsity issues. However, despite employing random sequence-based CL augmentation for inferring informative representation, DDGHM's intra-CL method does not explicitly consider the knowledge transfer and inter-domain correlation necessary for cross-domain modeling. C$^2$DSR \cite{C2CDR} is the most related CDR model, which develops an inter-domain contrastive infomax objective to improve the correlation between the domain-aware prototype representations and corresponding corrupted representations. 
However, its inter-CL objective over-maximizes the mutual information of the individual representations between the mixed and target domain, disregarding the potential similarity conflicts that may result in optimization collapse. Both coarse-grained similarity (refer to positive transfer) and fine-grained distinction (refer to negative transfer) should be considered in inter-domain CL.

Rather than solely relying on inter-CL between two domains, Tri-CDR incorporates all three domains and carefully model their triple relationships in contrastive tasks, ensuring that the optimization does not excessively collapse and get sufficient and accurate training. Precisely, Tri-CDR creatively proposes the coarse-gained similarity modeling and the fine-gained distinction modeling to comprehensively understand the triple correlations. The former captures the coarse-gained similarities between any two single-domain sequences of the same user to enable effective representation learning, while the latter maintains the robust cross-domain positive transfer through modeling the fine-gained distinction among triple sequences (see Sec. \ref{sec.ablation}). Moreover, as an effective and model-agnostic framework, Tri-CDR is able to bring the SOTA CL-based SR models (intra-domain CL) into the sequence modeling and achieves significant improvements in CDR by leveraging the combined interplay of both inter-CL and intra-CL strategies (see Sec. \ref{sec.universality}).

\subsubsection{\textbf{Comparison with Existing CDR Methods in Sequence Modeling Universality.}}

Behavior sequence modeling is essential in SR. Lots of novel techniques such as attention mechanisms \cite{CARCA, STOSA}, side information \cite{DIFSR,CARCA}, and contrastive learning \cite{CL4SRec,zhou2020s3} have been continuously proposed and verified to improve the performance of single-domain SR models. Amounts of CDR experiments have revealed that more recent and powerful single-domain SR models are able to outperform conventional CDSR algorithms \cite{DR-MTCDR, RecGURU, DisenCDR, DDGHM}, which is also observed in our experiments (see Sec. \ref{sec.results}).
However, most existing CDSR models commonly incorporate simple SR models as their sequence encoders to verify their CDR strategies, or design complicated and customized networks to learn multi-domain interactions tailored to their specific cross-domain settings, overlooking whether the proposed algorithm possess the flexibility to adapt to different (current or future-updated) strong sequence encoders.

Different from them, we argue that the universality of a CDSR algorithm with different sequence encoders is a crucial factor in guaranteeing the sustained effectiveness of the proposed framework, since real-world systems always prefer simple and universal methods. In this work, we try our best to enhance the universality and maintain the simplification of TCA and TCL without customized designs in the single-domain sequence modeling (see \ref{sec.universality} for detailed results). Therefore, Tri-CDR could leverage possible advancements in single-domain sequence modeling (including model evolution in the future), thereby extending the lifecycle of our proposed CDSR algorithms.

\subsubsection{\textbf{Complexity Analyses}}
In this section, we analyze the complexity of Tri-CDR and compare it with classical SR models (SASRec \cite{SASRec}, CL4SRec \cite{CL4SRec}) and CDSR algorithms (DASL \cite{DASL}, C$^2$DSR \cite{C2CDR}). 
The space complexity of the proposed Tri-CDR is largely determined by its sequence encoder. For example, apart from the mandatory space allocation for training three SASRec models (for source, mixed and target domains respectively), Tri-CDR (SASRec) only requires a limited number of MLPs (mentioned in Sec. \ref{par:TCA} and \ref{par:TCL}) to be additionally trained. That is, Tri-CDR does not introduce excessive trainable parameters, which renders the space complexity of Tri-CDR akin to its sequence encoder.

\begin{table}[t]
\caption{Statistics of three CDR datasets with six domains.}
\vspace{-0.3cm}
\label{tab:dataset}
\small
\begin{tabular}{|c|cc|cc|cc|}
\hline
\textbf{Dataset} &
  \multicolumn{2}{c|}{\begin{tabular}[c]{@{}c@{}}\textbf{Amazon Toy \& Game}\end{tabular}} &
  \multicolumn{2}{c|}{\begin{tabular}[c]{@{}c@{}}\textbf{Amazon Book \& Movie}\end{tabular}} &
  \multicolumn{2}{c|}{\begin{tabular}[c]{@{}c@{}}\textbf{Douban Book \& Music}\end{tabular}}\\ \hline
\textbf{Domain} & \multicolumn{1}{c|}{Toy} & \multicolumn{1}{c|}{Game} & \multicolumn{1}{c|}{Book} & \multicolumn{1}{c|}{Movie} & \multicolumn{1}{c|}{Book} & \multicolumn{1}{c|}{Music}     \\ \hline
\begin{tabular}[c]{@{}c@{}} \textbf{Users}\end{tabular} &
  \multicolumn{1}{c|}{7,996} &
  \multicolumn{1}{c|}{7,996} &
  \multicolumn{1}{c|}{28,531} &
  \multicolumn{1}{c|}{28,531} &
  \multicolumn{1}{c|}{4,580} &
  \multicolumn{1}{c|}{4,580} \\ \hline
\textbf{Items} & \multicolumn{1}{c|}{37,868}  & \multicolumn{1}{c|}{11,735} & \multicolumn{1}{c|}{239,042}   & \multicolumn{1}{c|}{38,185} & \multicolumn{1}{c|}{64,340}   & \multicolumn{1}{c|}{57,586}   \\ \hline
\textbf{Records} & \multicolumn{1}{c|}{114,487}  & \multicolumn{1}{c|}{82,871} & \multicolumn{1}{c|}{625,692}  & \multicolumn{1}{c|}{349,918} & \multicolumn{1}{c|}{224,117}  & \multicolumn{1}{c|}{278,855}  \\ \hline
\textbf{Density} & \multicolumn{1}{c|}{0.0378\%} & \multicolumn{1}{c|}{0.0883\%} & \multicolumn{1}{c|}{0.0092\%} & \multicolumn{1}{c|}{0.0321\%} & \multicolumn{1}{c|}{0.0761\%} & \multicolumn{1}{c|}{0.1057\%} \\ \hline
\end{tabular}
\vspace{-0.4cm}
\end{table}

As for the time complexity in model training, the base sequence encoders serve as the pivotal component of the CDSR tasks, and dominate its time complexity. Since different sequence encoders have different time complexity, we provide the time complexity of the used sequence encoders as follows: SASRec \cite{SASRec}: $\mathcal{O}((L^2+L)d|U|)$, CL4SRec \cite{CL4SRec}: $\mathcal{O}((L^2+L+B)d|U|)$, where $L$ denotes the length of the behavior sequence, $|U|$ denotes the number of users, $B$ is the size of mini-batch. Meanwhile, we also analyze the time complexity of the state-of-the-art CDSR algorithms which are armed with dual-attention mechanism (DASL \cite{DASL}), graph representation learning, and contrastive learning paradigm (C$^2$DSR \cite{C2CDR}). That is, the time complexity of DASL and C$^2$DSR is $\mathcal{O}((L^2+L)d|U|)$ and $\mathcal{O}((L^2+L+B+|\mathcal{R}^M|+|\mathcal{R}^T|)d|U|)$ respectively, where $|\mathcal{R}^M|$ and $|\mathcal{R}^T|$ denotes the number of nonzero entries in the Laplacian matrix in the mixed and target domain respectively. In contrast, we performed a comprehensive analysis of the time complexity of Tri-CDR on different sequence encoders, revealing that both Tri-CDR (SASRec) and Tri-CDR (CL4SRec) share the identical time complexity as $\mathcal{O}((L^2+L+B)d|U|)$. 

The aforementioned analyses indicate that Tri-CDR has asymptotic similar time complexity with CL4SRec in training. It is worth noting that Tri-CDR does not conduct TCL in the inference phase. Therefore, its online inference item complexity (which is the central metric considered in practical usages of recommendation models rather than the training time complexity) is comparable with DASL in magnitude. These computational complexities make Tri-CDR scalable on large cross-domain datasets. Therefore, the proposed Tri-CDR is able to employ cross-domain positive transfer and obtain high-quality sequence representations with only tolerant additional time cost and space cost.

\section{EXPERIMENTS}

In this section, we conduct extensive experiments to answer the following research questions: (RQ1): How does Tri-CDR perform against the state-of-the-art SR and CDSR baselines (see Sec. \ref{sec.results})? (RQ2): How do different components of Tri-CDR benefit its performance (see Sec. \ref{sec.ablation})? (RQ3): Is Tri-CDR still effective with other base sequence encoders (see Sec. \ref{sec.universality})? (RQ4): Does the introduction of additional domains lead to reasonable performance improvements? (see Sec. \ref{sec.domain_analysis})?(RQ5): How do some important hyper-parameters affect the performance of Tri-CDR (see Sec. \ref{sec.parameter})? (RQ6): How do the the coarse-grained similarity modeling and fine-grained distinction modeling contribute to the multi-domain representation learning in CDR(see Sec. \ref{sec.visualization})?

\subsection{Datasets}
To verify the effectiveness and universality of Tri-CDR, we conduct extensive experiments on six real-world CDR settings with four domains from Amazon \cite{Amazon} and two domains from Douban \cite{Douban}. Following \cite{DASL, DDGHM}, we select \emph{Amazon Book \& Movie}, \emph{Amazon Book \& Movie} and \emph{Douban Book \& Music} to form six CDR tasks. Amazon Toy \& Game, Amazon Book \& Movie and Douban Book \& Music include the review records from October 2000 to October 2018, from December 1996 to September 2018 and from July 2005 to December 2011 respectively. These three cross-domain datasets are pre-processed via the same method following classical CDR studies \cite{RecGURU, DASL}. Specifically, we build three user behavior sequences on the source, target, and mixed domains in chronological order. We set the last interacted item of each user as the test set and the penultimate interacted item as the valid set based on the Leave-one-out splitting method \cite{zhao2020revisiting, BERT4Rec}. We randomly select the users that have behaviors in both domains, filter out users having less than $3$ behaviors, and treat all rating records as interacted behaviors \cite{SIM, SSCDR}. The detailed statistics are shown in Table \ref{tab:dataset}.

\subsection{Baselines}
\label{sec.baseline}

In this section, we compare Tri-CDR with three representative SR models and five cross-domain SR models as follows:
\begin{itemize}[leftmargin=*]
   \item \textbf{GRU4Rec.} GRU4Rec \cite{GRU4Rec} is a classical session-based recommendation method encoding the item sequence via GRU.
   \item \textbf{SASRec.} SASRec \cite{SASRec} is a widely-used sequential recommendation model. It applies the self-attention mechanism to model behavior interactions in sequential recommendation.
   \item \textbf{CL4SRec.} CL4SRec \cite{CL4SRec} is the SOTA CL-based SR model, which adopts three sequence augmentation approaches to generate self-supervised signals via three intra-domain CL tasks.
\end{itemize}
Benefiting from the implementation of the self-attention mechanism, SASRec achieves a significant improvement relative to RNN-based algorithms in SR. For fair comparisons, we use the same sequence encoder of Tri-CDR (i.e., SASRec) in all cross-domain SR baselines (noted as \emph{Model+}). They also share the same features and historical behaviors:
\begin{itemize}[leftmargin=*]
   \item \textbf{SASRec(S+T).} SASRec(S+T) is a straightforward CDSR method based on SASRec. It first generates the source and target sequence representations respectively, then concatenates and feeds them into an MLP for the final prediction. 
    \item \textbf{DTCDR+.} DTCDR \cite{DTCDR} is a pioneer dual-target CDR method, which proposes an adaptable embedding sharing strategy to combine and share the user embeddings across domains based on the multi-task learning paradigm.
   \item \textbf{DDTCDR+.} DDTCDR \cite{DDTCDR} designs a latent orthogonal mapping function for extracting and transferring user preferences between two related domains while preserving cross-domain relations across different latent spaces in an iterative manner through a dual-transfer method.   
   \item \textbf{DASL+.} DASL \cite{DASL} constructs dual embeddings to extract the user's independent preferences and captures the user's cross-domain preference through a dual-attention learning mechanism. With its inherent structural superiority, DASL facilitates smooth incorporation of different sequence encoders. After replacing the sequence encoder from GRU to SASRec, DASL further achieves the promising performance, establishing itself as a strong baseline in CDSR.
    \item \textbf{C$^{2}$DSR+.} C$^{2}$DSR \cite{C2CDR} is the SOTA model in CDSR, which leverages a effective graph neural network to exploit the inter-domain co-occurrence collaborative relationship and proposes an contrastive infomax objective to capture and transfer the user's cross- domain preferences via the mutual information maximization mechanism. 
\end{itemize}


\begin{table*}[!htbp]
\caption{Results on cross-domain recommendation on Amazon platform. All improvements are significant (p<0.05 with paired t-tests).}
\vspace{-0.3cm}
\small
\label{tab:main_CDR}
\begin{tabular}{|c|c|c|c|c|c|c|c|c|c|c|}
\hline
Datasets &
  Algorithms &
  N@5 &
  N@10 &
  N@20 &
  N@50 &
  HR@5 &
  HR@10 &
  HR@20 &
  HR@50 &
  AUC \\ \hline
  
   &
  GRU4Rec &
  0.1305 &
  0.1551 &
  0.1783 &
  0.2172 &
  0.1860 &
  0.2624 &
  0.3549 &
  0.5533 &
  0.5613 \\
 &
  SASRec &
  0.1853 &
  0.2097 &
  0.2321 &
  0.2633 &
  0.2505 &
  0.3260 &
  0.4149 &
  0.5734 &
  0.5821 \\
 &
  CL4SRec &
  0.1898 &
  0.2143 &
  0.2359 &
  0.2673 &
  0.2577 &
  0.3338 &
  0.4199 &
  0.5795 &
  0.5828 \\ \cline{2-11}
 &
  SASRec(S+T) &
  0.1948 &
  0.2189 &
  0.2397 &
  0.2715 &
  0.2610 &
  0.3355 &
  0.4184 &
  0.5793 &
  0.5815 \\
 &
  DTCDR+ &
  0.1963 &
  0.2191 &
  0.2404 &
  0.2724 &
  0.2646 &
  0.3351 &
  0.4196 &
  0.5812 &
  0.5777 \\
 &
  DDTCDR+ &
  0.1959 &
  0.2185 &
  0.2404 &
  0.2723 &
  0.2617 &
  0.3316 &
  0.4186 &
  0.5800 &
  0.5768 \\
 &
  DASL+ &
  0.1976 &
  0.2206 &
  0.2418 &
  0.2740 &
  0.2624 &
  0.3338 &
  0.4183 &
  0.5810 &
  0.5788 \\ 
 &
  C$^{2}$DSR+ &
  0.1964 &
  0.2213 &
  0.2433 &
  0.2744 &
  0.2671 &
  0.3442 &
  0.4314 &
  0.5883 &
  0.5878 \\ \cline{2-11}
 &
  Tri-CDR(SASRec) &
  \textbf{0.2069} &
  \underline{0.2312} &
  \textbf{0.2528} &
  \underline{0.2832} &
  \textbf{0.2797} &
  \underline{0.3548} &
  \underline{0.4405} &
  \underline{0.5945} &
  \textbf{0.5913} \\
\multirow{-10}{*}{\begin{tabular}[c]{@{}c@{}}Game\\ ↓\\ Toy\end{tabular}} &
  Tri-CDR(CL4SRec) &
  \underline{0.2062} &
  \textbf{0.2313} &
  \underline{0.2525} &
  \textbf{0.2836} &
  \underline{0.2794} &
  \textbf{0.3576} &
  \textbf{0.4418} &
  \textbf{0.5989} &
  \underline{0.5895} \\ 
\hline

   &
  GRU4Rec &
  0.2682 &
  0.3053 &
  0.3366 &
  0.3726 &
  0.3697 &
  0.4839 &
  0.6079 &
  0.7894 &
  0.7510 \\
 &
  SASRec &
  0.3304 &
  0.3673 &
  0.3962 &
  0.4262 &
  0.4405 &
  0.5546 &
  0.6691 &
  0.8196 &
  0.7841 \\
 &
  CL4SRec &
  0.3317 &
  0.3687 &
  0.3969 &
  0.4264 &
  0.4434 &
  0.5579 &
  0.6695 &
  0.8178 &
  0.7835 \\ \cline{2-11}
 &
  SASRec(S+T) &
  0.3436 &
  0.3803 &
  0.4088 &
  0.4380 &
  0.4566 &
  0.5702 &
  0.6828 &
  0.8295 &
  0.7931 \\
 &
  DTCDR+ &
  0.3315 &
  0.3682 &
  0.3964 &
  0.4253 &
  0.4423 &
  0.5558 &
  0.6676 &
  0.8129 &
  0.7803 \\
 &
  DDTCDR+ &
  0.3355 &
  0.3708 &
  0.3989 &
  0.4282 &
  0.4473 &
  0.5566 &
  0.6677 &
  0.8146 &
  0.7832 \\
 &
  DASL+ &
  0.3368 &
  0.3734 &
  0.4010 &
  0.4299 &
  0.4482 &
  0.5613 &
  0.6706 &
  0.8155 &
  0.7819 \\ 
 &
  C$^{2}$DSR+ &
  0.3292 &
  0.3691 &
  0.3985 &
  0.4277 &
  0.4475 &
  0.5704 &
  0.6868 &
  0.8342 &
  0.7951 \\ \cline{2-11}
 &
  Tri-CDR(SASRec) &
  \underline{0.3514} &
  \underline{0.3892} &
  \underline{0.4182} &
  \underline{0.4458} &
  \underline{0.4684} &
  \textbf{0.5854} &
  \textbf{0.7000} &
  \textbf{0.8383} &
  \textbf{0.8015} \\
\multirow{-10}{*}{\begin{tabular}[c]{@{}c@{}}Toy\\ ↓\\ Game\end{tabular}} &
  Tri-CDR(CL4SRec) &
  \textbf{0.3562} &
  \textbf{0.3915} &
  \textbf{0.4205} &
  \textbf{0.4485} &
  \textbf{0.4712} &
  \underline{0.5806} &
  \underline{0.6954} &
  \underline{0.8357} &
  \underline{0.8004} \\ 
\hline
 &
  GRU4Rec &
  0.2381 &
  0.2695 &
  0.2994 &
  0.3421 &
  0.3199 &
  0.4173 &
  0.5362 &
  0.7521 &
  0.7163 \\
 &
  SASRec &
  0.2842 &
  0.3160 &
  0.3451 &
  0.3839 &
  0.3745 &
  0.4728 &
  0.5883 &
  0.7844 &
  0.7473 \\
 &
  CL4SRec &
  0.3013 &
  0.3340 &
  0.3627 &
  0.4001 &
  0.3925 &
  0.4937 &
  0.6078 &
  0.7971 &
  0.7586 \\ \cline{2-11}
 &
  SASRec(S+T) &
  0.2963 &
  0.3285 &
  0.3576 &
  0.3950 &
  0.3877 &
  0.4874 &
  0.6029 &
  0.7922 &
  0.7549 \\
 &
  DTCDR+ &
  0.3034 &
  0.3353 &
  0.3639 &
  0.4012 &
  0.3963 &
  0.4952 &
  0.6087 &
  0.7968 &
  0.7587 \\
 &
  DDTCDR+ &
  0.3018 &
  0.3339 &
  0.3629 &
  0.3995 &
  0.3939 &
  0.4935 &
  0.6081 &
  0.7937 &
  0.7569 \\
 &
  DASL+ &
  0.3027 &
  0.3346 &
  0.3635 &
  0.4004 &
  0.3946 &
  0.4933 &
  0.6080 &
  0.7948 &
  0.7576 \\ 
 &
  C$^{2}$DSR+ &
  0.3038 & 
  0.3359 &
  0.3651 &
  \underline{0.4015} &
  0.3974 &
  0.4966 &
  0.6126 &
  0.7967 &
  0.7592 \\ \cline{2-11}
 &
  Tri-CDR(SASRec) &
  \underline{0.3186} &
  \underline{0.3519} &
  \underline{0.3811} &
  \textbf{0.4171} &
  \textbf{0.4152} &
  \textbf{0.5182} &
  \textbf{0.6339} &
  \textbf{0.8156} &
  \textbf{0.7752} \\
\multirow{-10}{*}{\begin{tabular}[c]{@{}c@{}}Movie\\ ↓\\ Book\end{tabular}} &
  Tri-CDR(CL4SRec) &
  \textbf{0.3210} &
  \textbf{0.3533} &
  \textbf{0.3815} &
  \textbf{0.4171} &
  \underline{0.4140} &
  \underline{0.5142} &
  \underline{0.6264} &
  \underline{0.8063} &
  \underline{0.7693} \\ 
\hline
  
   &
  GRU4Rec &
  0.4123 &
  0.4477 &
  0.4735 &
  0.5002 &
  0.5338 &
  0.6432 &
  0.7448 &
  0.8795 &
  0.8401 \\
 &
  SASRec &
  0.4492 &
  0.4843 &
  0.5096 &
  0.5343 &
  0.5712 &
  0.6795 &
  0.7796 &
  0.9034 &
  0.8629 \\
 &
  CL4SRec &
  0.4578 &
  0.4930 &
  0.5177 &
  0.5417 &
  0.5812 &
  0.6897 &
  0.7872 &
  0.9078 &
  0.8677 \\ \cline{2-11}
 &
  SASRec(S+T) &
  0.4594 &
  0.4944 &
  0.5191 &
  0.5430 &
  0.5834 &
  0.6913 &
  0.7888 &
  0.9090 &
  0.8687 \\
 &
  DTCDR+ &
  0.4508 &
  0.4864 &
  0.5115 &
  0.5357 &
  0.5779 &
  0.6878 &
  0.7869 &
  0.9082 &
  0.8671 \\
 &
  DDTCDR+ &
  0.4590 &
  0.4938 &
  0.5187 &
  0.5424 &
  0.5827 &
  0.6898 &
  0.7883 &
  0.9070 &
  0.8674 \\
 &
  DASL+ &
  0.4591 &
  0.4939 &
  0.5189 &
  0.5425 &
  0.5827 &
  0.6899 &
  0.7885 &
  0.9072 &
  0.8676 \\
 &
  C$^{2}$DSR+ &
  0.4587 &
  0.4945 &
  0.5189 &
  0.5424 &
  0.5869 &
  \underline{0.6973} &
  \underline{0.7936} &
  \underline{0.9117} &
  \underline{0.8713} \\ \cline{2-11}
 &
  Tri-CDR(SASRec) &
  \textbf{0.4669} &
  \textbf{0.5015} &
  \textbf{0.5258} &
  \textbf{0.5491} &
  \textbf{0.5933} &
  \textbf{0.6999} &
  \textbf{0.7960} &
  \textbf{0.9128} &
  \textbf{0.8729} \\ 
\multirow{-10}{*}{\begin{tabular}[c]{@{}c@{}}Book\\ ↓\\ Movie\end{tabular}} &
  Tri-CDR(CL4SRec) &
  \underline{0.4667} &
  \underline{0.5005} &
  \underline{0.5244} &
  \underline{0.5478} &
  \underline{0.5923} &
  0.6966 &
  0.7914 &
  0.9089 &
  0.8694 \\ \hline
\end{tabular}
\vspace{-0.4cm}
\end{table*}

\subsection{Experimental Settings}
We implement the above methods using PyTorch with python 3.8.10. For fair comparisons, we take Adam as the optimizing method and the learning rate is set as $0.0005$. We initialize model parameters randomly using the Xavier method. The batch size and the dimension of embedding size are set as $120$ and $64$. We adopt the same maximum length of sequence for each model, which is $200$ on all datasets. We conduct a grid search to select hyper-parameters. Specifically, we select the $\lambda_{CSM}$ and $\lambda_{FDM}$ in \{0.1, 0.5, 1, 4, 10\}. For the Amazon dataset, which exhibits significant disparities in sparsity between the two domains, we define the ratio between $\lambda_{1}$, $\lambda_{2}$ and $\lambda_{3}$ as 1:1:1 for the sparser Amazon Toy and Amazon Book and 100:1:1 and 1000:1:1 for the denser Amazon Movie and Amazon Game respectively. In contrast, we define the ratio between $\lambda_{1}$, $\lambda_{2}$ and $\lambda_{3}$ as 1:1:1 for the Douban dataset with similar sparsity of two domains. The temperature coefficient $\tau$ is set to be $0.1$. According to the average natural distribution of the behavior data in Amazon dataset, we set $\gamma$ to $4.0$ in the sparser Amazon Toy and Amazon Book, $0.5$ in the denser Amazon Game and Amazon Movie. Regarding for the Douban dataset, we define the $\gamma$ as $0.5$ and $0.05$ for Douban Book and Douban Music respectively. We conduct five runs with different random seeds and report the average results of all models.

\begin{table}[t]
\caption{Results on cross-domain recommendation on Douban platform. All improvements are significant (p<0.05 with paired t-tests).}
\vspace{-0.3cm}
\label{tab:main_douban}
\small
\begin{tabular}{|c|c|c|c|c|c|c|c|c|c|c|}
\hline
\textbf{Datasets} &
  \textbf{Algorithms} &
  \textbf{N@5} &
  \textbf{N@10} &
  \textbf{N@20} &
  \textbf{N@50} &
  \textbf{HR@5} &
  \textbf{HR@10} &
  \textbf{HR@20} &
  \textbf{HR@50} &
  \textbf{AUC} \\ \hline
\multirow{10}{*}{\begin{tabular}[c]{@{}c@{}}Music\\ ↓\\ Book\end{tabular}} & GRU4Rec          & 0.4643 & 0.4944 & 0.5175 & 0.5441 & 0.5815 & 0.6746 & 0.7659 & 0.8998 & 0.8583 \\ 
                                           & SASRec           & 0.5336 & 0.5632 & 0.5838 & 0.6048 & 0.6503 & 0.7414 & 0.8231 & 0.9281 & 0.8912 \\ 
                                           & CL4SRec          & 0.5211 & 0.5514 & 0.5717 & 0.5927 & 0.6454 & 0.7386 & 0.8187 & 0.9244 & 0.8876 \\ \cline{2-11} 
                                           & DTCDR+           & 0.5413 & 0.5699 & 0.5898 & 0.6100 & 0.6600 & 0.7482 & 0.8269 & 0.9284 & 0.8927 \\  
                                           & SASRec(S+T)      & 0.5518 & 0.5813 & 0.6022 & 0.6206 & 0.6687 & 0.7598 & 0.8424 & \underline{0.9348} & 0.8993 \\ 
                                           & DDTCDR+          & 0.5374 & 0.5671 & 0.5876 & 0.6074 & 0.6544 & 0.7461 & 0.8270 & 0.9268 & 0.8918 \\ 
                                           & DASL+            & 0.5501 & 0.5806 & 0.6015 & 0.6199 & 0.6664 & 0.7602 & 0.8427 & \textbf{0.9350} & 0.8993 \\ 
                                           & C2DSR            & 0.5523 & 0.5823 & 0.6024 & 0.6205 & 0.6717 & \underline{0.7641} & \underline{0.8436} & 0.9346 & \underline{0.8996} \\ \cline{2-11} 
                                           & Tri-CDR(SASRec)  & \textbf{0.5625} & \textbf{0.5924} & \textbf{0.6118} & \textbf{0.6287} & \textbf{0.6815} & \textbf{0.7737} & \textbf{0.8503} & \textbf{0.9350} & \textbf{0.9028} \\ 
                                           & Tri-CDR(CL4SRec) & \underline{0.5571} & \underline{0.5860} & \underline{0.6043} & \underline{0.6232} & \underline{0.6743} & 0.7635 &     0.8358 & 0.9301 & 0.8973 \\ \hline     
                                           
\multirow{10}{*}{\begin{tabular}[c]{@{}c@{}}Book\\ ↓\\ Music\end{tabular}} & GRU4Rec          & 0.5030 & 0.5338 & 0.5547 & 0.5757 & 0.6310 & 0.7258 & 0.8085 & 0.9136 & 0.8774 \\ 
                                           & SASRec           & 0.5674 & 0.5973 & 0.6160 & 0.6331 & 0.6900 & 0.7818 & 0.8559 & 0.9413 & 0.9079 \\ 
                                           & CL4SRec          & 0.5516 & 0.5815 & 0.5997 & 0.6174 & 0.6842 & 0.7761 & 0.8480 & 0.9371 & 0.9031 \\ \cline{2-11} 
                                           & DTCDR+           & 0.5689 & 0.5973 & 0.6163 & 0.6331 & 0.6937 & 0.7813 & 0.8561 & 0.9401 & 0.9082 \\ 
                                           & SASRec(S+T)      & 0.5771 & 0.6062 & 0.6248 & 0.6409 & 0.7015 & 0.7913 & 0.8646 & \underline{0.9449} & 0.9124 \\ 
                                           & DDTCDR+          & 0.5720 & 0.6005 & 0.6191 & 0.6360 & 0.6935 & 0.7814 & 0.8548 & 0.9398 & 0.9076 \\ 
                                           & DASL+            & 0.5788 & 0.6089 & 0.6272 & 0.6430 & 0.7012 & 0.7936 & 0.8658 & 0.9447 & 0.9130 \\ 
                                           & C2DSR            & 0.5806 & 0.6101 & 0.6284 & 0.6440 & 0.7043 & \underline{0.7949} & \underline{0.8670} & \underline{0.9449} & \underline{0.9136} \\ \cline{2-11} 
                                           & Tri-CDR(SASRec)  & \textbf{0.5891} & \textbf{0.6177} & \textbf{0.6354} & \textbf{0.6505} & \textbf{0.7140} & \textbf{0.8019} & \textbf{0.8717} & \textbf{0.9472} & \textbf{0.9165} \\ 
                                           & Tri-CDR(CL4SRec) & \underline{0.5826} & \underline{0.6112} & \underline{0.6291} & \underline{0.6447} & \underline{0.7051} & 0.7934 &     0.8640 & 0.9421 & 0.9124 \\ \hline
                                           
\end{tabular}
\vspace{-0.4cm}
\end{table}

\vspace{-0.2cm}
\subsection{Performance Comparison on Cross-domain Sequential Recommendation (RQ1)}
\label{sec.results}
We conduct our experiments on six CDR tasks, adopting three typical evaluation metrics including NDCG@k (N@k), Hit Rate@k (HR@k), and AUC with different $k=5,10,20,50$. Following \cite{SASRec}, we randomly sample 99 negative items for each positive instance in testing phase. Table \ref{tab:main_CDR} and Table \ref{tab:main_douban} shows the results on the Amazon platform and Douban platform respectively, and we can observe that:

(1) Tri-CDR achieves the best performances on all metrics and datasets compared to all baselines, with the significance level $p<0.05$. The NDCG@10 improvements of Tri-CDR over the best baseline are $4.52\%$/$2.95\%$/$5.18\%$/$1.42\%$ on Amazon Toy/Game/Book/Movie, and the HR@10 improvements are $3.89\%$/$2.63\%$/$4.25\%$/$0.37\%$ on four Amazon datasets consistently. Similarly, the NDCG@10 improvements of Tri-CDR over the best baseline are $1.73\%$/$1.25\%$ on Douban Book/Music, and the HR@10 improvements are $1.26\%$/$0.88\%$.
It indicates that our triple sequence learning is beneficial in CDR. Moreover, it also demonstrates that Tri-CDR can well model the correlations among users' triple behavior sequences, and successfully capture useful information related to the target-domain prediction from all domains.

(2) Tri-CDR outperforms all CDR baselines that also consider multi-domain behaviors. It confirms the significance of (a) explicit triple sequence learning with the mixed behavior sequence that contains the user's global interests, and (b) our TCL and TCA that could better model triple correlations among three domains and combine them into user representations.
Comparing with SOTA CDSR models we can know that, DDTCDR and DASL focus on the dual transfer between source and target domains. They perform worse than Tri-CDR due to the lack of triple sequence learning. C$^2$DSR also conducts contrastive infomax objectives to improve global-local dual correlations. However, it does not consider the fine-grained distinctions in domain correlation modeling and triple cross-domain attention in multi-domain combination. Sometimes the infomax loss is not well functioned, which may be caused by its unreal negative sequences.
Tri-CDR also outperforms different ablation versions besides baselines as shown in Sec. \ref{sec.ablation}.

(3) Comparing the improvements among different domains, we find that Tri-CDR is more beneficial on Game$\rightarrow$ Toy(sparse) and Movie$\rightarrow$Book(sparse) settings. It not only confirms the effectiveness of Tri-CDR on different data distributions, but also reflects that Tri-CDR functions well on relatively sparser target domains, where the positive knowledge transfer from the source domain should bring in more essential information as supplements, implying the practical usage of Tri-CDR. 

(4) To confirm its universality with different application platforms, we have also evaluated Tri-CDR on the Douban dataset (Book \& Music) besides Amazon datasets. 
In combination with the statistics of CDR datasets in Table \ref{tab:main_douban}, we discover that: (a) consistent with the conventional CDSR algorithms, Tri-CDR functions well on relatively sparser target domains by transferring the positive knowledge from the denser source domain. (b) In contrast to the cross-domain settings on the Amazon platform, Tri-CDR shows relatively similar improvements in the two cross-domain settings on Douban. This may be attributed to the fact that users have longer average interactions on Douban dataset (48.93 and 60.89 for Book and Music respectively).
(c) The experimental results also demonstrate that some dual-modeling CDSR methods can obtain promising performance on the Douban dataset. However, Tri-CDR outperforms these algorithms on most metrics, which provides further evidence of Tri-CDR's significant role in alleviating cross-domain negative transfer and accurately modeling the correlations among the triple domains. We further conduct a universality analysis on Tri-CDR adopted with different base sequence encoders in Sec. \ref{sec.universality}.

\begin{table*}[t]
\caption{Results on ablation study of Tri-CDR(SASRec).}
\vspace{-0.3cm}
\label{tab:ablation}
\small
\begin{tabular}{|c|l|c|c|c|c|c|c|c|c|c|}
\hline
Datasets &
  Models &
  N@5 &
  N@10 &
  N@20 &
  N@50 &
  HR@5 &
  HR@10 &
  HR@20 &
  HR@50 &
  AUC \\ \hline
\multirow{4}{*}{\begin{tabular}[c]{@{}c@{}}Game\\ ↓\\ Toy\end{tabular}} &
  Tri-CDR w/o TCA and TCL &
  0.2005 &
  0.2239 &
  0.2470 &
  0.2778 &
  0.2720 &
  0.3446 &
  0.4364 &
  0.5925 &
  0.5833 \\
 &
  Tri-CDR w/o TCA &
  0.2022 &
  0.2274 &
  0.2502 &
  0.2807 &
  0.2745 &
  0.3522 &
  \textbf{0.4426} &
  0.5971 &
  \textbf{0.5925} \\
 &
  Tri-CDR w/o FDM &
  0.2030 &
  0.2280 &
  0.2503 &
  0.2814 &
  0.2757 &
  0.3530 &
  0.4416 &
  \textbf{0.5991} &
  0.5915 \\
 &
  Tri-CDR &
  \textbf{0.2069} &
  \textbf{0.2312} &
  \textbf{0.2528} &
  \textbf{0.2832} &
  \textbf{0.2797} &
  \textbf{0.3548} &
  0.4405 &
  0.5945 &
  0.5913 \\ \hline
  
\multirow{4}{*}{\begin{tabular}[c]{@{}c@{}}Toy\\ ↓\\ Game\end{tabular}} &
  Tri-CDR w/o TCA and TCL &
  0.3258 &
  0.3636 &
  0.3937 &
  0.4234 &
  0.4408 &
  0.5578 &
  0.6770 &
  0.8262 &
  0.7866 \\
 &
  Tri-CDR w/o TCA &
  0.3509 &
  0.3871 &
  0.4152 &
  0.4430 &
  \textbf{0.4698} &
  0.5816 &
  0.6928 &
  0.8322 &
  0.7949 \\
 &
  Tri-CDR w/o FDM &
  0.3485 &
  0.3861 &
  0.4153 &
  0.4429 &
  0.4673 &
  0.5837 &
  0.6990 &
  0.8376 &
  \textbf{0.8017} \\
 &
  Tri-CDR &
  \textbf{0.3514} &
  \textbf{0.3892} &
  \textbf{0.4182} &
  \textbf{0.4458} &
  0.4684 &
  \textbf{0.5854} &
  \textbf{0.7000} &
  \textbf{0.8383} &
  0.8015 \\ \hline
\multirow{4}{*}{\begin{tabular}[c]{@{}c@{}}Movie\\ ↓\\ Book\end{tabular}} &
  Tri-CDR w/o TCA and TCL &
  0.2959 &
  0.3279 &
  0.3577 &
  0.3952 &
  0.3874 &
  0.4866 &
  0.6047 &
  0.7945 &
  0.7554 \\
 &
  Tri-CDR w/o TCA &
  0.3085 &
  0.3399 &
  0.3679 &
  0.4048 &
  0.3995 &
  0.4967 &
  0.6076 &
  0.7941 &
  0.7589 \\
 &
  Tri-CDR w/o FDM &
  0.3088 &
  0.3415 &
  0.3706 &
  0.4071 &
  0.4044 &
  0.5057 &
  0.6211 &
  0.8055 &
  0.7667 \\
 &
  Tri-CDR &
  \textbf{0.3186} &
  \textbf{0.3519} &
  \textbf{0.3811} &
  \textbf{0.4171} &
  \textbf{0.4152} &
  \textbf{0.5182} &
  \textbf{0.6339} &
  \textbf{0.8156} &
  \textbf{0.7752} \\ \hline
\multirow{4}{*}{\begin{tabular}[c]{@{}c@{}}Book\\ ↓\\ Movie\end{tabular}} &
  Tri-CDR w/o TCA and TCL &
  0.4379 &
  0.4754 &
  0.5008 &
  0.5250 &
  0.5706 &
  0.6860 &
  0.7865 &
  0.9077 &
  0.8658 \\
 &
  Tri-CDR w/o TCA &
  0.4611 &
  0.4953 &
  0.5196 &
  0.5433 &
  0.5849 &
  0.6902 &
  0.7864 &
  0.9054 &
  0.8661 \\
 &
  Tri-CDR w/o FDM &
  0.4505 &
  0.4858 &
  0.5106 &
  0.5344 &
  0.5830 &
  0.6916 &
  0.7895 &
  0.9088 &
  0.8683 \\
 &
  Tri-CDR &
  \textbf{0.4669} &
  \textbf{0.5015} &
  \textbf{0.5258} &
  \textbf{0.5491} &
  \textbf{0.5933} &
  \textbf{0.6999} &
  \textbf{0.7960} &
  \textbf{0.9128} &
  \textbf{0.8729} \\ \hline

\multirow{4}{*}{\begin{tabular}[c]{@{}c@{}}Book\\ ↓\\ Music\end{tabular}} &
    Tri-CDR w/o TCA and TCL & 0.5798 & 0.6092 & 0.6269 & 0.6427 & 0.7067 & 0.7970  & 0.8670  & 0.9461 & 0.9139 \\
    &
    Tri-CDR w/o TCA & 0.5868 & 0.6160  & 0.6339 & 0.6491 & 0.7099 & 0.7997 & 0.8703 & 0.9465 & 0.9152 \\
    &
    Tri-CDR w/o FDM & 0.5836 & 0.6138 & 0.6313 & 0.6468 & 0.7084 & 0.8014 & 0.8704 & \textbf{0.9480}  & 0.9155 \\
    &
    Tri-CDR         & \textbf{0.5891} & \textbf{0.6177} & \textbf{0.6354} & \textbf{0.6505} & \textbf{0.7140}  & \textbf{0.8019} & \textbf{0.8717} & 0.9472 & \textbf{0.9165} \\ \hline
\multirow{4}{*}{\begin{tabular}[c]{@{}c@{}}Music\\ ↓\\ Book\end{tabular}} &
    Tri-CDR w/o TCA and TCL & 0.5532 & 0.5833 & 0.6031 & 0.6209 & 0.6721 & 0.7649 & 0.8433 & 0.9324 & 0.8991 \\
    & 
    Tri-CDR w/o TCA & 0.5576 & 0.5881 & 0.6073 & 0.6254 & 0.6754 & 0.7693 & 0.8452 & 0.9361 & 0.901  \\
    & 
    Tri-CDR w/o FDM & 0.5595 & 0.5902 & 0.6100   & 0.6271 & 0.6777 & 0.7728 & \textbf{0.8509} & \textbf{0.9368} & \textbf{0.9028} \\
    & 
    Tri-CDR         & \textbf{0.5625} & \textbf{0.5924} & \textbf{0.6118} & \textbf{0.6287} & \textbf{0.6815} & \textbf{0.7737} & 0.8503 & 0.9350 & \textbf{0.9028} \\ \hline

\end{tabular}
\vspace{-0.4cm}
\end{table*}

\subsection{Ablation Study (RQ2)}
\label{sec.ablation}
In this section, we aim to find whether different components are effective in Tri-CDR. Thus we compare Tri-CDR with Tri-CDR w/o TCA\&TCL, Tri-CDR w/o TCA and Tri-CDR w/o FDM to verify the benefits of TCA, TCL and FDM. In general, most components' improvements are significant (the average error range $\leq0.003$). 
In Table \ref{tab:ablation} we have:

(1) Tri-CDR w/o TCA performs significantly better than Tri-CDR w/o TCA\&TCL, verifying the effectiveness of TCL. TCL takes full advantage of CL's alignment and uniformity \cite{wang2020understanding,yu2022graph} and extends it to triple domains, maximizing the multi-domain mutual information while remaining necessary preference diversity in knowledge transfer.

(2) Tri-CDR further improves the results of Tri-CDR w/o TCA. Our TCA highlights the information related to the target-domain preference learned from the current target-domain historical behaviors as well as the user's global interests learned from mixed behaviors via cross-domain attention, which enables more positive knowledge transfer and is beneficial for CDR.

(3) Comparing Tri-CDR with and without FDM, we further demonstrate that the fine-grained distinction modeling in TCL is indispensable. FDM keeps the domain diversity and significantly brings in additional $1.60\%$/$0.78\%$ average NDCG@10/HR@10 improvements on six datasets. Besides, it not only helps Tri-CDR to learn better multi-domain sequence representations, but also makes the model training more stable with different parameters.

\begin{figure*}[t]
\centering
\includegraphics[width=0.98\textwidth]{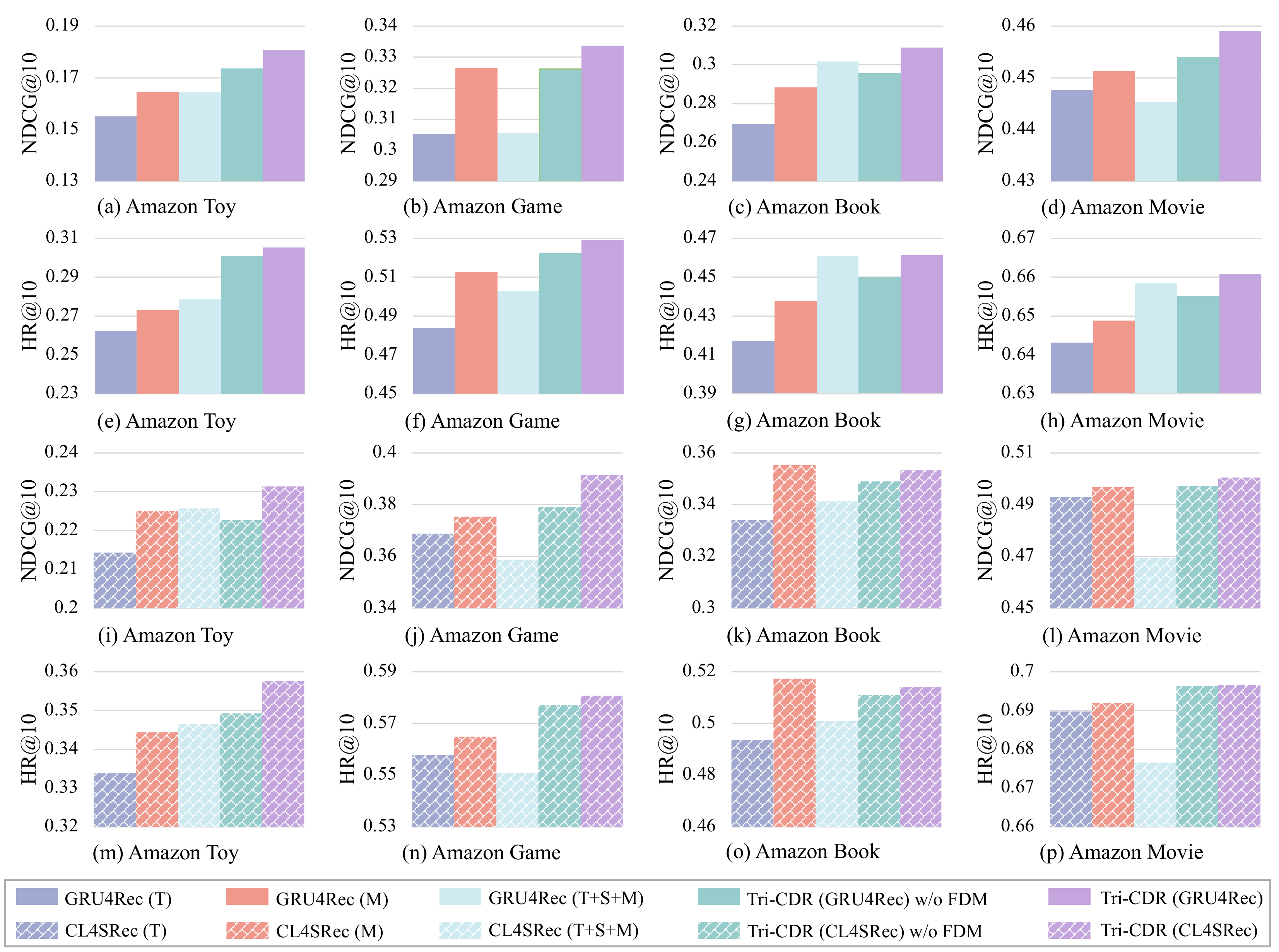}
\vspace{-0.2cm}
\caption{Universality analysis on Tri-CDR. We show the results of different versions of Tri-CDR on GRU4Rec and CL4SRec.}
\vspace{-0.4cm}
\label{fig:universality}
\end{figure*}

\subsection{Universality of Tri-CDR (RQ3)}
\label{sec.universality}

Tri-CDR is a model-agnostic framework. We further evaluate its universality on Amazon datasets based on GRU4Rec and CL4SRec. Fig. \ref{fig:universality} illustrates Tri-CDR models on GRU4Rec and CL4SRec, and we can find that:

(1) In general, Tri-CDR still achieves consistent and significant improvements over different ablation versions on all datasets with both GRU4Rec and CL4SRec, which confirms its universality when adopted with different sequential models and even other CL tasks. The improvements are consistent with other metrics.

(2) Comparing with different Tri-CDR's ablation versions, we reconfirm that (a) the mixed behavior sequence is informative, while directly combining source, target, and mixed sequences may also bring in noises, and (b) the proposed TCA and TCL with FDM are effective to make full use of information of three domains in CDR.

(3) We should highlight that CL4SRec also conducts intra-domain CL tasks based on some sequence augmentations. The improvement brought by Tri-CDR implies that our inter-domain CL could cooperate well with various intra-domain CL. We notice that CL4SRec(M) has comparable or even better results over Tri-CDR on the Book domain after careful parameter selections. It is because that the intra-domain CL of CL4SRec on the mixed sequence (containing multi-domain behaviors) works as certain inter-domain CL tasks. Nevertheless, Tri-CDR still achieves the best results in general.

\begin{table*}[t]
\centering
\caption{Results on different domain utilization of Tri-CDR (SASRec).}
\vspace{-0.2cm}
\label{tab:domain_analysis}
\small
\begin{tabular}{|c|l|c|c|c|c|c|c|c|c|c|}
\hline
Datasets &
  Models &
  N@5 &
  N@10 &
  N@20 &
  N@50 &
  HR@5 &
  HR@10 &
  HR@20 &
  HR@50 &
  AUC \\ \hline
\multirow{6}{*}{\begin{tabular}[c]{@{}c@{}}Game\\ ↓\\ Toy\end{tabular}} &
  SASRec (T) &
  0.1853 &
  0.2097 &
  0.2321 &
  0.2633 &
  0.2505 &
  0.3260 &
  0.4149 &
  0.5734 &
  0.5821 \\
 &
  SASRec (M) &
  0.1953 &
  0.2200 &
  0.2417 &
  0.2744 &
  0.2654 &
  0.3420 &
  0.4280 &
  0.5941 &
  \textbf{0.6017} \\
 &
  SASRec (S+T) &
  0.1948 &
  0.2189 &
  0.2397 &
  0.2715 &
  0.2610 &
  0.3355 &
  0.4184 &
  0.5793 &
  0.5815 \\
 &
  SASRec (M+T) &
  0.1965 &
  0.2210 &
  0.2439 &
  0.2745 &
  0.2675 &
  0.3434 &
  0.4347 &
  0.5893 &
  0.5908 \\
 &
  SASRec (S+T+M) &
  0.2005 &
  0.2239 &
  0.2470 &
  0.2778 &
  0.2720 &
  0.3446 &
  0.4364 &
  0.5925 &
  0.5833 \\
 &
  Tri-CDR &
  \textbf{0.2069} &
  \textbf{0.2312} &
  \textbf{0.2528} &
  \textbf{0.2832} &
  \textbf{0.2797} &
  \textbf{0.3548} &
  \textbf{0.4405} &
  \textbf{0.5945} &
  0.5913 \\ \hline
\multirow{6}{*}{\begin{tabular}[c]{@{}c@{}}Toy\\ ↓\\ Game\end{tabular}} &
  SASRec (T) &
  0.3304 &
  0.3673 &
  0.3962 &
  0.4262 &
  0.4405 &
  0.5546 &
  0.6691 &
  0.8196 &
  0.7841 \\
 &
  SASRec (M) &
  0.3408 &
  0.3780 &
  0.4064 &
  0.4359 &
  0.4553 &
  0.5706 &
  0.6826 &
  0.8310 &
  0.7966 \\
 &
  SASRec (S+T) &
  0.3436 &
  0.3803 &
  0.4088 &
  0.4380 &
  0.4566 &
  0.5702 &
  0.6828 &
  0.8295 &
  0.7931 \\
&
  SASRec (M+T) &
  0.3191 &
  0.3576 &
  0.3880 &
  0.4180 &
  0.4359 &
  0.5549 &
  0.6752 &
  0.8256 &
  0.7878 \\
 &
  SASRec (S+T+M) &
  0.3305 &
  0.3687 &
  0.3988 &
  0.4281 &
  0.4463 &
  0.5648 &
  0.6836 &
  0.8310 &
  0.7904 \\
 &
  Tri-CDR &
  \textbf{0.3514} &
  \textbf{0.3892} &
  \textbf{0.4182} &
  \textbf{0.4458} &
  \textbf{0.4684} &
  \textbf{0.5854} &
  \textbf{0.7000} &
  \textbf{0.8383} &
  \textbf{0.8015} \\ \hline
\multirow{6}{*}{\begin{tabular}[c]{@{}c@{}}Movie\\ ↓\\ Book\end{tabular}} &
  SASRec (T) &
  0.2842 &
  0.3160 &
  0.3451 &
  0.3839 &
  0.3745 &
  0.4728 &
  0.5883 &
  0.7844 &
  0.7473 \\
 &
  SASRec (M) &
  0.2852 &
  0.3183 &
  0.3479 &
  0.3868 &
  0.3757 &
  0.4783 &
  0.5958 &
  0.7925 &
  0.7533 \\
 &
  SASRec (S+T) &
  0.2963 &
  0.3285 &
  0.3576 &
  0.3950 &
  0.3877 &
  0.4874 &
  0.6029 &
  0.7922 &
  0.7549 \\
 &
  SASRec (M+T) &
  0.3000 &
  0.3326 &
  0.3617 &
  0.3985 &
  0.3922 &
  0.4930 &
  0.6083 &
  0.7948 &
  0.7573 \\
 &
  SASRec (S+T+M) &
  0.2959 &
  0.3279 &
  0.3577 &
  0.3952 &
  0.3874 &
  0.4866 &
  0.6047 &
  0.7945 &
  0.7554 \\
 &
  Tri-CDR &
  \textbf{0.3186} &
  \textbf{0.3519} &
  \textbf{0.3811} &
  \textbf{0.4171} &
  \textbf{0.4152} &
  \textbf{0.5182} &
  \textbf{0.6339} &
  \textbf{0.8156} &
  \textbf{0.7752} \\ \hline
\multirow{6}{*}{\begin{tabular}[c]{@{}c@{}}Book\\ ↓\\ Movie\end{tabular}} &
  SASRec (T) &
  0.4492 &
  0.4843 &
  0.5096 &
  0.5343 &
  0.5712 &
  0.6795 &
  0.7796 &
  0.9034 &
  0.8629 \\
 &
  SASRec (M) &
  0.4639 &
  0.4988 &
  0.5234 &
  0.5470 &
  0.5873 &
  0.6949 &
  0.7924 &
  0.9103 &
  0.8704 \\
 &
  SASRec (S+T) &
  0.4594 &
  0.4944 &
  0.5191 &
  0.5430 &
  0.5834 &
  0.6913 &
  0.7888 &
  0.9090 &
  0.8687 \\
 &
  SASRec (M+T) &
  0.4486 &
  0.4849 &
  0.5100 &
  0.5336 &
  0.5809 &
  0.6927 &
  0.7918 &
  0.9105 &
  0.8693 \\
 &
  SASRec (S+T+M) &
  0.4469 &
  0.4814 &
  0.5062 &
  0.5303 &
  0.5749 &
  0.6814 &
  0.7792 &
  0.9002 &
  0.8611 \\
 &
  Tri-CDR &
  \textbf{0.4669} &
  \textbf{0.5015} &
  \textbf{0.5258} &
  \textbf{0.5491} &
  \textbf{0.5933} &
  \textbf{0.6999} &
  \textbf{0.7960} &
  \textbf{0.9128} &
  \textbf{0.8729} \\ \hline
\end{tabular}
\vspace{-0.3cm}
\end{table*}

\subsection{Analyses on the domain utilization of Tri-CDR (RQ4)}
\label{sec.domain_analysis}

To inspect the effectiveness of different domains on Tri-CDR, we use S, T, and M to represent using source, target, and mixed sequences respectively, and compare Tri-CDR with five combinations. It is worth noting that SASRec (T) and SASRec (M) denote the single-domain SR models with the user's target behavior sequence and mixed behavior sequence. On the other hand, SASRec (S+T), SASRec (M+T), and SASRec (S+M+T) refer to the cross-domain SR models based on the user's source and target domain \cite{piNet, DTCDR,DDTCDR,DASL}, mixed and target domain \cite{DDGHM, C2CDR}, and mixed, source and target domain (the proposed CDSR setting), respectively. From Table. \ref{tab:domain_analysis}, we can observe that: 

(a) Comparing the first two versions, SASRec (M) significantly outperforms SASRec (T) on all metrics and datasets. It is reasonable since the mixed sequence is the complete user chronological behavior sequence from both the source and target domains. The sequential encoder on the mixed behavior sequence can be viewed as the cross-domain sequence encoder to some extent, yielding cross-domain information gain. 

(b) Dual-modeling CDR methods (SASRec (S+T) and SASRec (M+T)) do not always achieve consistent improvement over the single-domain SR methods, and the simply triple-modeling CDR method (SASRec (S+T+M)) may even lead to further performance deterioration in some cross-domain settings. It further reinforces that both source and mixed sequences encompass cross-domain knowledge and noise information simultaneously and that simple dual- and triple-modeling strategies may be insufficient to accurately distinguish and model the complex correlations and confounding knowledge in multiple sequences.

(c) Comparing SASRec (S+T+M) and Tri-CDR on four cross-domain settings, we further demonstrate the effectiveness of the proposed TCA and TCL, which is primarily due to the fact that TCA highlights the users' target-domain preferences and comprehensive interests from the cross-domain knowledge transfer, and TCL precisely models the correlation among triple behavior sequences.

\begin{figure}[t]
\centering
\vspace{-0.2cm}
\includegraphics[width=0.99\columnwidth]{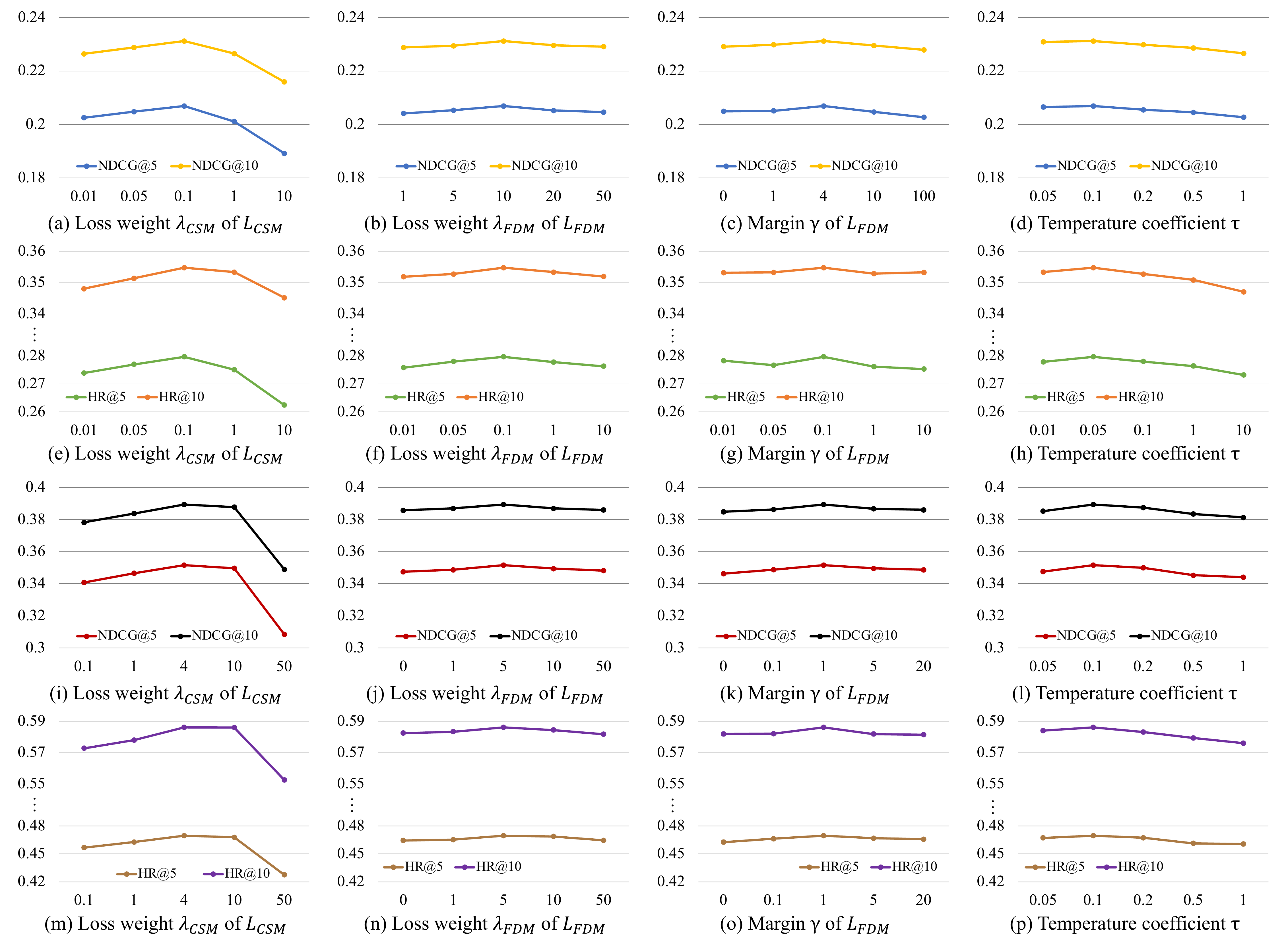}
\caption{Parameter analyses on loss weight $\lambda_{CSM}$ of $\mathcal{L}_{CSM}$, $\lambda_{FDM}$ of $\mathcal{L}_{FDM}$, margin $\gamma$ of $\mathcal{L}_{FDM}$ and temperature coefficient $\tau$. (a)-(h) shows the results of Game$\rightarrow$Toy, and (i)-(p) shows the results of Toy$\rightarrow$Game.}
\vspace{-0.4cm}
\label{fig:parameter_experiment}
\end{figure}

\subsection{Parameter Sensitivity Analyses (RQ5)}
\label{sec.parameter}

\subsubsection{Analyses on CL loss weights $\lambda_{CSM}$, $\lambda_{FDM}$, margin $\gamma$ and temperature coefficient $\tau$}

In Fig. \ref{fig:parameter_experiment}, we conduct four parameter analyses on the Game$\rightarrow$Toy setting (the first two rows) and the Toy$\rightarrow$Game setting (the last two rows) to investigate the performance trends with different loss weight $\lambda_{CSM}$ of $\mathcal{L}_{CSM}$, loss weights $\lambda_{FDM}$ of $\mathcal{L}_{FDM}$, margin $\gamma$ of $\mathcal{L}_{FDM}$ and temperature coefficient $\tau$ in Eq. (\ref{eq.l_csm}) and Eq. (\ref{eq.L_FD}). We observe that:
(1) Tri-CDR's performance first increases and then decreases as $\lambda_{CSM}$ and $\tau$ gets larger. According to the behavior distribution characteristics, the loss weight $\lambda_{FDM}$ of $\mathcal{L}_{FDM}$ differs for different cross-domain settings. $\lambda_{CSM}=0.1 / 4.0$ achieves the best performance on Game$\rightarrow$Toy and Toy$\rightarrow$Game settings, respectively. In contrast, Tri-CDR is insensitive to the temperature coefficient $\tau$, so we set it as $0.1$ for all CDSR settings with the purpose of simplifying the hyper-parameter tuning and obtaining the promising performance. 
(2) Too smaller $\lambda_{FDM}$ may weaken the power of FDM in TCL, while too larger $\lambda_{FDM}$ may also disturb the triple correlation learning in CSM. It is also natural that Tri-CDR is less sensitive to fine-grained $\lambda_{FDM}$ compared to coarse-grained $\lambda_{CSM}$. The optimal performance of Tri-CDR is observed when $\lambda_{FDM}$ is set to $10.0$ and $5.0$ on Game$\rightarrow$Toy and Toy$\rightarrow$Game settings.
(3) $\gamma=0$ indicates that the model only wants the source-mixed distance to be smaller than the source-target distance. When $\gamma=100$, Eq. (\ref{eq.L_FD}) is always active to broaden the fine-grained cross-domain distance gap. Tri-CDR performs relatively poor under the extreme values of both sides, indicating the importance of an appropriate margin in FDM ($4.0$ and $1.0$ for the Game$\rightarrow$Toy and Toy$\rightarrow$Game settings).

\begin{table}[t]
\caption{Parameter analyses on the ratio among loss weights $\lambda_1$, $\lambda_2$ and $\lambda_3$}
\vspace{-0.3cm}
\label{tab:cl_ratio}
\small
\resizebox{\textwidth}{!}{
\begin{tabular}{|c|c|c|c|c|c|c|c|c|c|c|c|}
\hline
\textbf{Datasets} & \textbf{Models} & \textbf{\begin{tabular}[c]{@{}c@{}}Ratio among \\ $\lambda_1$, $\lambda_2$ and $\lambda_3$\end{tabular}} & \textbf{N@5} & \textbf{N@10} & \textbf{N@20} & \textbf{N@50} & \textbf{HR@5} & \textbf{HR@10} & \textbf{HR@20} & \textbf{HR@50} & \textbf{AUC} \\ \hline
\multirow{10}{*}{\begin{tabular}[c]{@{}c@{}}Game\\ ↓\\ Toy\end{tabular}} & \multirow{5}{*}{\begin{tabular}[c]{@{}c@{}}Tri-CDR\\ w/o FDM\end{tabular}} 
& 1:1:0.1 & 0.2036 & 0.2272 & 0.2496 & 0.2807 & 0.2770 & 0.3501 & 0.4387 & 0.5954 & 0.5867 \\ \cline{3-12} 
&  & 1:1:1 & 0.2030 & 0.2280 & 0.2503 & 0.2814 & 0.2757 & 0.3530 & 0.4416 & 0.5991 & 0.5915 \\ \cline{3-12} 
&  & 10:1:1 & 0.2016 & 0.2258 & 0.2481 & 0.2791 & 0.2730 & 0.3479 & 0.4362 & 0.5934 & 0.5861 \\ \cline{3-12} 
&  & 100:1:1 & 0.1992 & 0.2232 & 0.2453 & 0.2763 & 0.2682 & 0.3428 & 0.4307 & 0.5877 & 0.5835 \\ \cline{3-12} 
&  & 1000:1:1 & 0.1978 & 0.2223 & 0.2444 & 0.2754 & 0.2666 & 0.3425 & 0.4304 & 0.5873 & 0.5833 \\ \cline{2-12} 
& \multirow{5}{*}{Tri-CDR} 
& 1:1:0.1 & 0.2072 & 0.2320 & 0.2538 & 0.2840 & 0.2803 & 0.3568 & 0.4433 & 0.5961 & 0.5916 \\ \cline{3-12} 
&  & 1:1:1 & 0.2069 & 0.2312 & 0.2528 & 0.2832 & 0.2797 & 0.3548 & 0.4405 & 0.5945 & 0.5913 \\ \cline{3-12} 
&  & 10:1:1 & 0.2033 & 0.2275 & 0.2490 & 0.2793 & 0.2731 & 0.3479 & 0.4336 & 0.5868 & 0.5867 \\ \cline{3-12} 
&  & 100:1:1 & 0.2008 & 0.2246 & 0.2457 & 0.2769 & 0.2689 & 0.3429 & 0.4266 & 0.5844 & 0.5831 \\ \cline{3-12} 
&  & 1000:1:1 & 0.2003 & 0.2244 & 0.2461 & 0.2768 & 0.2676 & 0.3421 & 0.4282 & 0.5842 & 0.5832 \\ \hline
\multirow{10}{*}{\begin{tabular}[c]{@{}c@{}}Toy\\ ↓\\ Game\end{tabular}} & \multirow{5}{*}{\begin{tabular}[c]{@{}c@{}}Tri-CDR\\ w/o FDM\end{tabular}} 
& 1:1:1 & 0.3276 & 0.3664 & 0.3971 & 0.4268 & 0.4449 & 0.5647 & 0.6866 & 0.8354 & 0.7959 \\ \cline{3-12} 
&  & 10:1:1 & 0.3336 & 0.3722 & 0.4016 & 0.4309 & 0.4498 & 0.5688 & 0.6852 & 0.8324 & 0.7930 \\ \cline{3-12} 
&  & 100:1:1 & 0.3423 & 0.3801 & 0.4092 & 0.4381 & 0.4566 & 0.5737 & 0.6887 & 0.8337 & 0.7949 \\ \cline{3-12} 
&  & 1000:1:1 & 0.3485 & 0.3861 & 0.4153 & 0.4429 & 0.4673 & 0.5837 & 0.6990 & 0.8376 & 0.8017 \\ \cline{3-12} 
&  & 1000:1:0.1 & 0.3476 & 0.3845 & 0.4143 & 0.4424 & 0.4655 & 0.5795 & 0.6975 & 0.8385 & 0.8010 \\ \cline{2-12} 
& \multirow{5}{*}{Tri-CDR} & 1:1:1 & 0.3326 & 0.3708 & 0.4002 & 0.4302 & 0.4490 & 0.5671 & 0.6834 & 0.8339 & 0.7948 \\ \cline{3-12} 
&  & 10:1:1 & 0.3437 & 0.3809 & 0.4101 & 0.4384 & 0.4602 & 0.5752 & 0.6907 & 0.8328 & 0.7951 \\ \cline{3-12} 
&  & 100:1:1 & 0.3463 & 0.3842 & 0.4127 & 0.4406 & 0.4631 & 0.5807 & 0.6934 & 0.8330 & 0.7957 \\ \cline{3-12} 
&  & 1000:1:1 & 0.3514 & 0.3892 & 0.4182 & 0.4458 & 0.4684 & 0.5854 & 0.7000 & 0.8383 & 0.8015 \\ \cline{3-12} 
&  & 1000:1:0.1 & 0.3503 & 0.3877 & 0.4170 & 0.4444 & 0.4687 & 0.5843 & 0.6998 & 0.8375 & 0.8013 \\ \hline
\end{tabular}}
\vspace{-0.4cm}
\end{table}

\subsubsection{Analyses on loss weights $\lambda_1$, $\lambda_2$, and $\lambda_3$ in CSM}
\label{subsec.loss_weight}
We implement a series of experiments to investigate the effect of different ratios among $\lambda_1$, $\lambda_2$, and $\lambda_3$ of $\mathcal{L}_{CSM}$ in Tri-CDR. Table \ref{tab:cl_ratio} shows the results of Tri-CDR(SASRec) with or without FDM on the Game$\rightarrow$Toy and Toy$\rightarrow$Game settings. We observe that:
(1) for Tri-CDR w/o FDM, larger loss weights of $\lambda_1$ lead to better performance on the Toy$\rightarrow$Game setting (the relatively denser target domains). It is natural since the correlations between the source and mixed behavior sequences should be more highlighted to ensure learning informative source sequence representations from the sparser source behavior sequences. For the Game$\rightarrow$Toy setting (the sparser target domains), we find that the loss weight ratio of 1:1:1 already achieves the best performance.
(2) For Tri-CDR, we also observe the same trend on the Toy$\rightarrow$Game setting that the performances increase as $\lambda_1$ becomes larger. Moreover, regardless of the ratio among $\lambda_1$, $\lambda_2$, and $\lambda_3$, Tri-CDR shows consistent improvements compared to Tri-CDR w/o FDM. Enhanced by FDM, the results are relatively satisfactory even with imperfect loss weights (average significant improvements of $1.17\%$ and $0.51\%$ on NDCG@10 and HR@10 respectively). It demonstrates that the proposed FDM helps to improve the effectiveness as well as the robustness of triple contrastive learning, making it less sensitive to different loss weights in CSM and more practical in real-world scenarios.
(3) In comparing the different performances of Tri-CDR with different ratios between $\lambda_2$, and $\lambda_3$ in two cross-domain settings, we notice that Tri-CDR exhibited further improvements after reducing $\lambda_3$ (loss weight of $\mathcal{L}_CSM$ between source and target domains) on the Game$\rightarrow$Toy setting. This improvement is interpretable as its underlying principle aligns with certain assumptions of FDM. Specifically, Tri-CDR precisely controls the similarity between the source and target domains by constraining the hyper-parameters, thereby modeling the fine-grained distinction among triple sequences. To mitigate the complexity of model training while preserving its scalability, we maintain the balanced ratio of $\lambda_2$, and $\lambda_3$ as 1:1 when presenting the performance of Tri-CDR. Experimental results also indicate that Tri-CDR achieves significant improvements over the current state-of-the-art baseline, even without fine-grained tuning of the ratio between $\lambda_2$, and $\lambda_3$.

\subsection{Estimation of User Preference Modeling in Tri-CDR (RQ6)}
\label{sec.visualization}
\begin{figure}[!t]
\centering
\includegraphics[width=0.9\columnwidth]{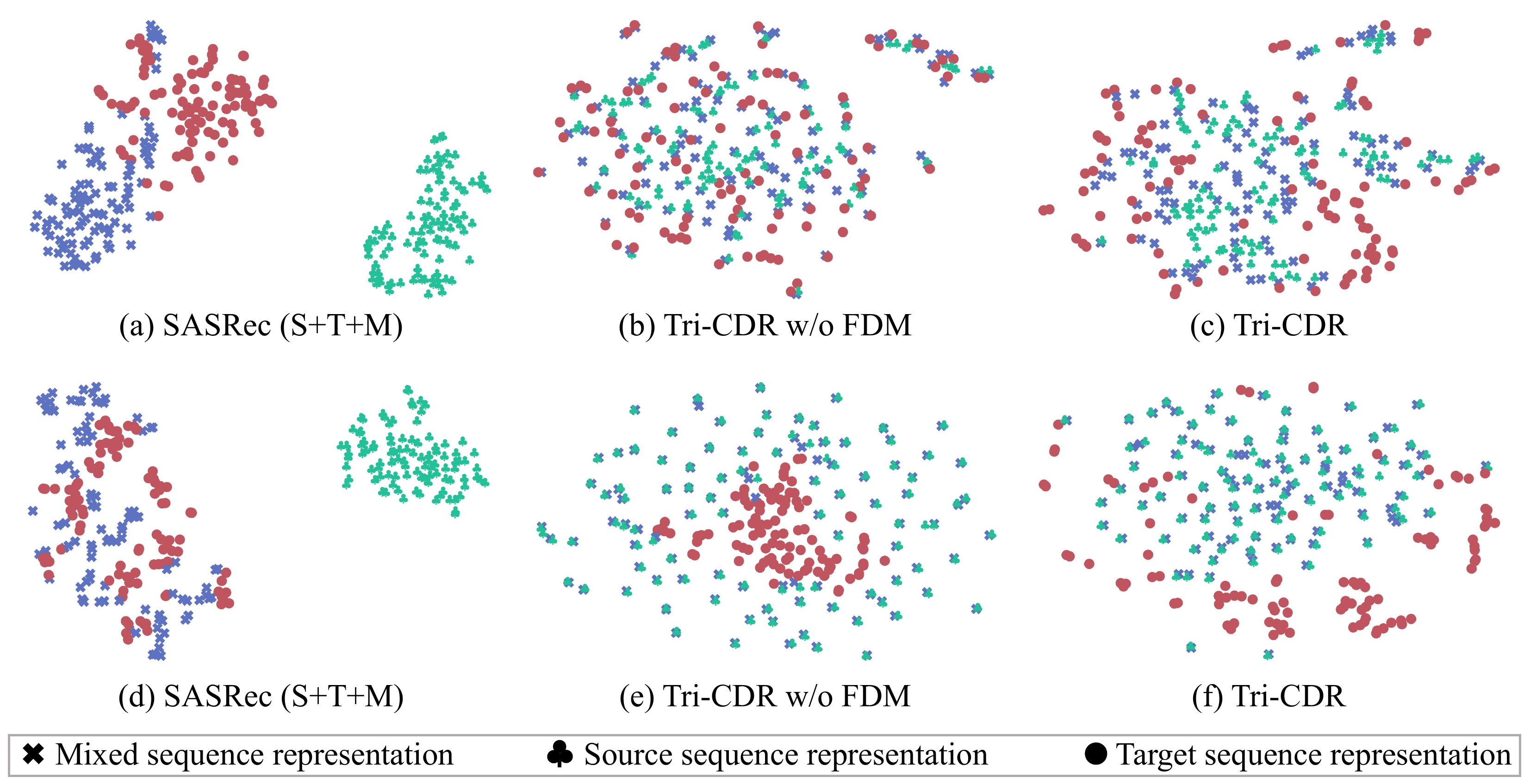}
\vspace{-0.2cm}
\caption{Visualization of different Tri-CDR versions among triple domains, where the blue cross, green clubs, and red circle denote the mixed sequence representation, source sequence representation and target sequence representation respectively.}
\vspace{-0.4cm}
\label{fig:tsne_visualization_domain}
\end{figure}

\begin{figure}[!t]
\centering
\includegraphics[width=0.9\columnwidth]{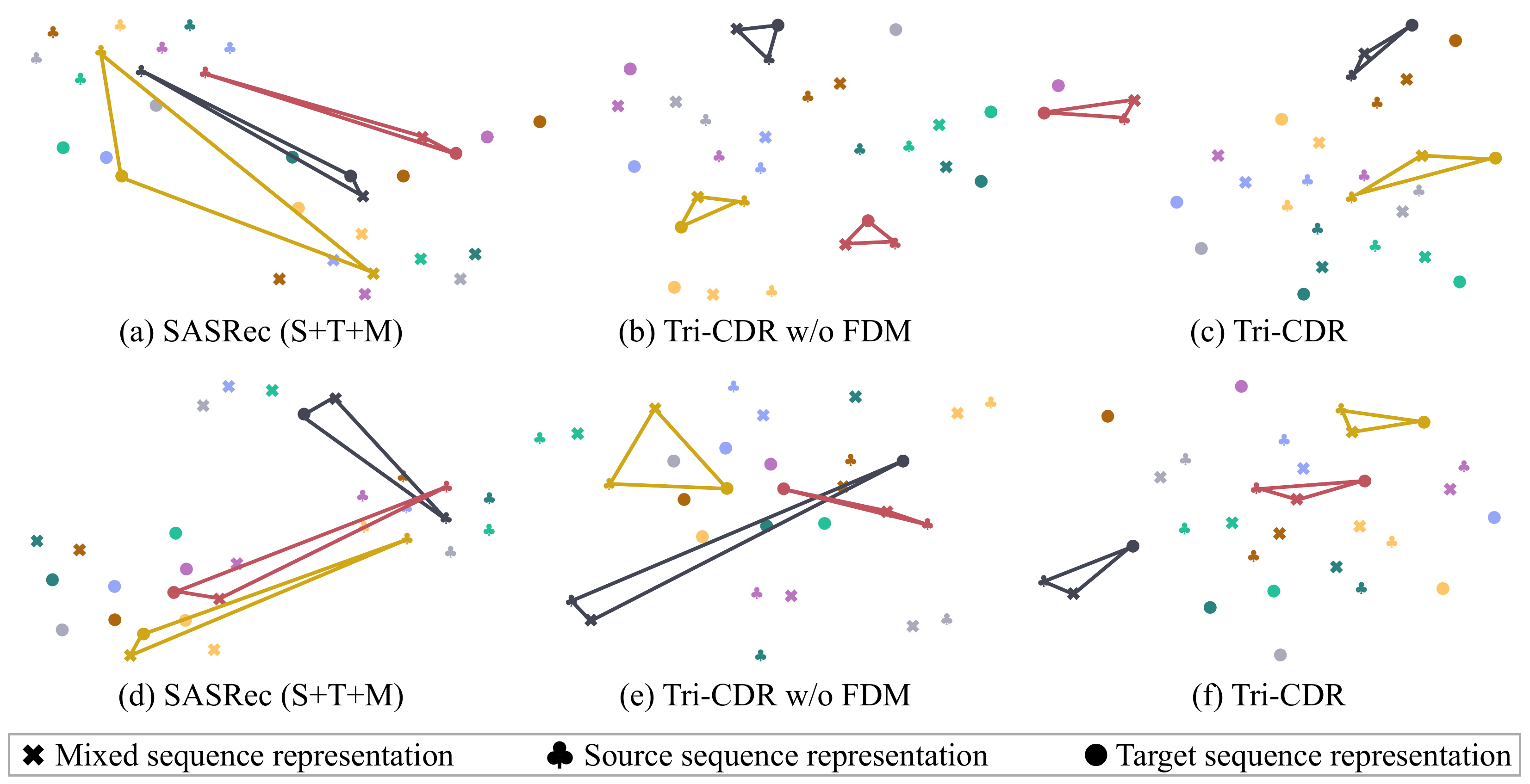}
\vspace{-0.2cm}
\caption{Visualization of different Tri-CDR versions from the perspective of different users. We employ color differentials to distinguish different users while utilizing shapes to differentiate the same user's sequential representations across triple domains.}
\vspace{-0.4cm}
\label{fig:tsne_visualization_user}
\end{figure}

To intuitively show the impacts of both similarity and distinction modeling in TCL, we show the visualization of randomly selected multi-domain sequence representations in SASRec (S+T+M), Tri-CDR w/o FDM, and Tri-CDR via t-Distributed Stochastic Neighbor Embedding (t-SNE) \cite{tsne}. Fig. \ref{fig:tsne_visualization_domain} and Fig. \ref{fig:tsne_visualization_user} illustrate the distributions of three domains and different users via different colors and shapes, respectively. In each figure, the first row refers to the visualization on the Game$\rightarrow$Toy setting, while the second row represents the visualization on the Toy$\rightarrow$Game setting.

\subsubsection{Visualization of the overall distribution across triple domains.}
We depict the overall distribution of the randomly selected 100 users' sequential representations across different domains via t-SNE in two cross-domain settings, as illustrated in Fig. \ref{fig:tsne_visualization_domain}. We observe that: 
(1) As shown in Fig. \ref{fig:tsne_visualization_domain} (a) and (d), most users' multi-domain sequence representations are naturally clustered via their domains rather than their users in SASRec (S+T+M). 
(2) Comparing Fig. \ref{fig:tsne_visualization_domain} (b) and (e) with Fig. \ref{fig:tsne_visualization_domain} (a) and (d), the multi-domain sequence representations are converted from domain-based clustering to the user-based, which indicates that the coarse-grained similarity modeling in TCL does make triple sequence representations of a user to be similar. 
(3) In contrast to Fig. \ref{fig:tsne_visualization_domain} (e), the target domain sequence representations in Fig. \ref{fig:tsne_visualization_domain} (f) do not aggregate in the approximate location but rather distributed around user-specific cross-domain information. This indicates that FDM enhances the discriminability of the target domain sequence representations in coarse-grained comparisons, enabling precise modeling of the users' target interests.

\subsubsection{Visualization of the independent distribution among different users.}
In order to investigate the independent distribution of different domain sequence representations among different users, we randomly select 10 users and visualized the aforementioned representations with t-SNE. The observations are as follows: 
(1) Similar to Fig. \ref{fig:tsne_visualization_domain}(a) and (d), the distances among the same user's triple sequence representations are relatively large in \ref{fig:tsne_visualization_user}(a) and (d). This leads to a failure in building adequate cross-domain correlations, and thus cannot make full use of additional source/mixed domains’ information.
(2) In Fig. \ref{fig:tsne_visualization_user}(b) and (e), irrespective of whether adjust the ratio hyper-parameters or not, the association among triple sequence representations of the same user is consistently improved with CSM. This proves the effectiveness of the information gain derived from coarse-grained similarity modeling.
(3) In Fig. \ref{fig:tsne_visualization_user}(b), some users' multi-domain sequence representations are too close to form small acute triangles. Too homogeneous multi-domain representations may weaken the additional information gains from source/mixed sequences, which will harm the positive knowledge transfer. In contrast, armed with FDM, these users in Fig. \ref{fig:tsne_visualization_user}(c) have more distinguishable sequence representations forming obtuse triangles, and thus Tri-CDR could achieve better results.
(4) Despite the employment of sufficient tuning ratios among $\lambda_1$, $\lambda_2$, and $\lambda_3$ in CSM, Fig. \ref{fig:tsne_visualization_user}(e) still exhibits unexpected deviations in the triple correlation. In contrast, the visualization of triangles in Fig. \ref{fig:tsne_visualization_user}(f) provides tangible evidence of the practical significance of FDM.

\section{CONCLUSION}
In this work, we propose a model-agnostic Triple sequence learning for Cross-Domain Recommendation (Tri-CDR) framework. Conventional CDR methods mainly focus on modeling the dual-relations between the source and target domains or the mixed and target domains, failing to explore the triple correlation among the source, mixed, and target domains. Tri-CDR conducts a triple cross-domain attention method to highlight useful information and accelerate positive knowledge transfer and enables a more accurate multi-domain sequence representation learning strategy via both the coarse-grained similarity modeling and fine-grained distinction modeling. The extensive evaluation and analyses on six benchmark cross-domain settings demonstrate that Tri-CDR is able to precisely model the similarity while preserving the information diversity among triple domains, which reveals the underlying principles of its effectiveness and universality. We believe that the triple sequence learning paradigm will provide a solid foundation for researchers and practitioners to explore new directions in cross-domain recommendation.

In the future, we will continue to explore the correlations among the source, target, and natural mixed behavior sequences, as well as more sophisticated modeling on their representation learning and multi-domain aggregation. We will also enhance Tri-CDR's capability by incorporating more modality information of item contents as semantic bridges in multi-domain recommendation.

\begin{acks}
To Robert, for the bagels and explaining CMYK and color spaces.
\end{acks}
\bibliographystyle{ACM-Reference-Format}
\bibliography{sample-base}

\end{document}